\documentclass[useAMS,usenatbib]{mn2e}
\usepackage{graphicx,natbib,amsmath,subfig}
\input journals.def
\title[Asymmetric supernova in hierarchical systems and application to J1903]
  {Asymmetric supernova in hierarchical multiple star systems and application to J1903+0327}
\author[J.T. Pijloo, D.P. Caputo and S.F. Portegies Zwart]
  {J.T.~Pijloo$^1$\thanks{The first and second authors contributed equally to the work.}, 
  D.P.~Caputo$^1$\footnotemark[1], 
  and S.F.~Portegies Zwart$^1$ \\ 
  $^1$Leiden Observatory, Leiden University, P.O. Box 9513, 2300 RA Leiden, the Netherlands\\}
\date{Released 2011 Xxxxx XX}

\pagerange{\pageref{firstpage}--\pageref{lastpage}} \pubyear{2011}

\def\LaTeX{L\kern-.36em\raise.3ex\hbox{a}\kern-.15em
    T\kern-.1667em\lower.7ex\hbox{E}\kern-.125emX}

\begin{document}

\label{firstpage}

\maketitle

\begin{abstract}
We develop a method to analyze the effect of an asymmetric supernova
on hierarchical multiple star systems and we present analytical
formulas to calculate orbital parameters for surviving binaries or
hierarchical triples and runaway velocities for their dissociating
equivalents. The effect of an asymmetric supernova on the orbital
parameters of a binary system has been studied to great extent
(e.g. \citealt{1983ApJ...267..322H,1996ApJ...471..352K,1998A&A...330.1047T}),
but this effect on higher multiplicity hierarchical systems has not
been explored before. With our method, the supernova effect can be
computed by reducing the hierarchical multiple to an effective binary
by means of recursively replacing the inner binary by an effective
star at the center of mass of that binary. We apply our method to a
hierarchical triple system similar to the progenitor of PSR J1903+0327
suggested by \citet{2011ApJ...734...55P}. We confirm their earlier finding that
PSR J1903+0327 could have evolved from a hierarchical triple that became
unstable and ejected the secondary star of the inner binary.  Furthermore,
if such as system did evolve via this mechanism the most probable configuration
would be a small supernova kick velocity, an inner binary with a large
semi-major axis, and the fraction of mass accreted onto the neutron star to
the mass lost by the secondary would most likely be between 0.35 and 0.5
\end{abstract}

\begin{keywords}
 stars: supernovae: general -- binaries: general -- methods: analytical -- methods: numerical -- pulsars: individual: J1903+0327.
\end{keywords}

\section{Introduction}\label{sec:Introduction}
Asymmetric supernovae (SNe) in binary and hierarchical multiple star systems
form a crucial phase in the formation of stellar systems containing a
compact stellar remnant - neutron star or black hole. In previous studies
of SNe in binaries, two effects of the SN have been considered: (1) sudden
mass loss and (2) a random kick velocity imparted on the compact remnant 
of the star undergoing the SN. The combined effect which changes the orbital 
parameters causes the binary to dissociate in the majority of the cases.

The study of binaries surviving a supernova (SN) explosion of one of
its components was first performed by \citet{1961BAN....15..265B} and
\citet{1961BAN....15..291B}, assuming a symmetric SN (i.e. only mass
loss). The necessity of asymmetry in the SN, resulting in the kick
velocity, was first suggested by \citet{1970SvA....13..562S}. The
statistical study on pulsar scale heights by
\citet{1970ApJ...160..979G} firmly supported the asymmetric SN model
and to date the adding of the kick velocity to the newly born neutron
star (or black hole) is a commonly excepted mechanism
\citep{1997ApJ...483..399V}. Both the type of explosion mechanism and
whether the exploding star is in a binary system are found to
influence the effect of the kick velocity (see
e.g. \citealt{2004ApJ...612.1044P}), but the exact physical process
underlying the production of kicks remains unclear. The analysis of
the effect of asymmetric supernovae on binaries has been sufficient to
explain most of the observed post-SN stellar systems, and little to no
effort has gone into studying the effect on hierarchical multiple star
systems.

Millisecond pulsar (MSP) J1903+0327 (spin period $\simeq$ 2.15 ms), first observed by
\citet{2008Sci...320.1309C} and later, in more detail, by
\citet{2011MNRAS.412.2763F}, is part of what may be the first observed
MSP binary to have evolved from a hierarchical triple progenitor.  MSP
J1903+0327 is orbited by a main sequence star in a wide (orbital
period $\simeq$ 95.2 days) and eccentric (eccentricity $e$
$\simeq$ 0.44) orbit. Based on these observables it seems impossible
that this binary (hereafter J1903+0327) formed via the traditional
mechanism in a binary progenitor
\citep{2008Sci...320.1309C}. \citet{2011ApJ...734...55P} proposed that
the progenitor system was a binary accompanied by a third and least
massive main-sequence star in a wider orbit about this binary. During
the low-mass X-ray binary (LMXB) phase of the inner binary, the orbit
of the LMXB expanded due to mass transfer from the evolving inner
companion (donor) star to the neutron star, which was formed in the
SN. This eventually caused the triple to become dynamically unstable
and to eject the inner companion resulting in the observed system
J1903+0327.

 J1903+0327 is not a unique case, however: there is a
significant number of systems like the progenitor of J1903+0327 as
suggested in \citet{2011ApJ...734...55P} and similar hierarchical
stellar systems of higher multiplicity. The Multiple Star Catalog
lists 602 triples, 93 quadruples, 22 quintuples, 9 sextuples and 2
septuples \citep{1997A&AS..124...75T} of which 90 systems contain at least one star
with a mass $M \geq 10$ M$_\odot$. Each of these multiples will
eventually experience a core-collapse SN of the most massive
star. After the SN these systems are either fully dissociated,
dissociate into lower multiplicity multiple star systems, or survive
the SN.

 We begin the study of the effect of an asymmetric SN on
hierarchical multiple star systems by first readdressing the SN effect
on a binary and subsequently treating the effect in a hierarchical
triple. We show that a hierarchical triple can effectively be regarded
as a binary system comprised of the center of mass of the inner binary
and the tertiary star. The effect of a SN on a hierarchical triple
system, now reduced to an effective binary, can be calculated using
the prescription for a SN in binary. We ultimately generalize this
effective binary method to hierarchical multiple star systems of
arbitrary multiplicity. In the second part of the paper we perform
Monte Carlo simulations of a hierarchical triple star system similar
to the progenitor of J1903+0327 suggested in
\citet{2011ApJ...734...55P} to determine the (stable) survival rates,
and evaluate whether such a formation route is plausible.

\section{Calculation of post-SN parameters}\label{sec:Parameter_calculations}
\subsection{Binary systems}\label{subsec:Binary}
We consider a binary system of stars with mass, position and velocity
for the primary and secondary star, given by
($m_{1,0}$,$\mathbf{r_1}$,$\mathbf{v_{1,0}}$) and
($m_2$,$\mathbf{r_2}$,$\mathbf{v_{2,0}}$) respectively\footnote{The
  contingent suffix 1, 2, etc. indicates which star we are considering
  (e.g. 1 for the primary). The contingent suffix 0 denotes the pre-SN
  state and when it is absent, it either refers to the post-SN state
  or the absence indicates that there is no difference in the pre- and
  post-SN states of that parameter.}, in which the primary undergoes a
SN. The binary system is uniquely described by the semi-major axis,
$a_0$, eccentricity, $e_0$, and true anomaly, $\theta_0$. The
separation distance is $\mathbf{r_0}$. We assume that the SN is
instantaneous, meaning an instantaneous removal of mass of the
primary, no SN-shell impact on the companion (secondary) star,
and the orbital motion during this mass loss phase is neglected,
i.e. $\mathbf{r} = \mathbf{r_0}$ and $\mathbf{v_2} =
\mathbf{v_{2,0}}$.

\begin{figure}
  \centering
  \includegraphics[width=\linewidth]{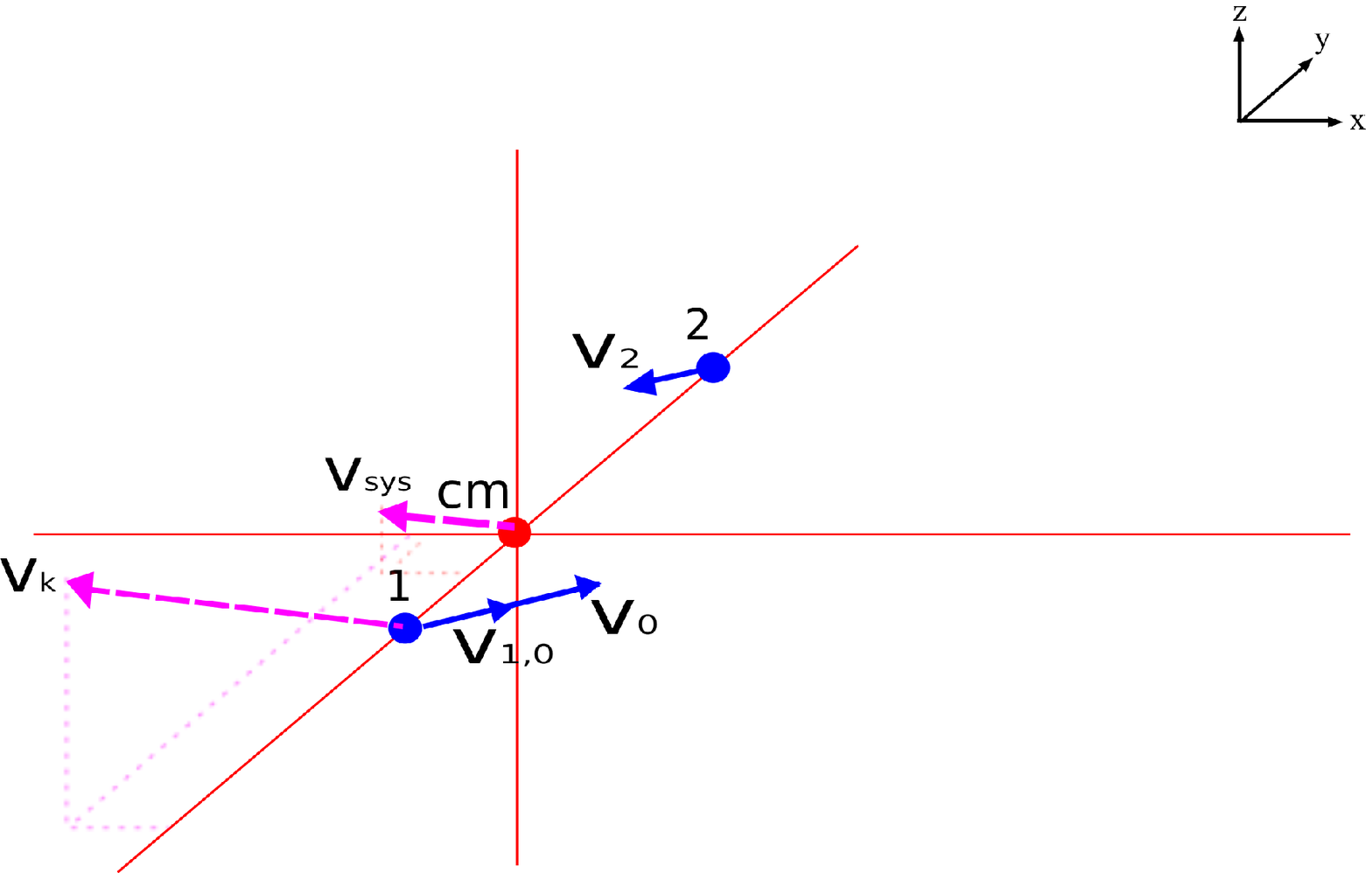}
  ~ a. The cm coordinate system in the cm reference frame for a binary
  system before the SN (at $t = 0$).
  \includegraphics[width=\linewidth]{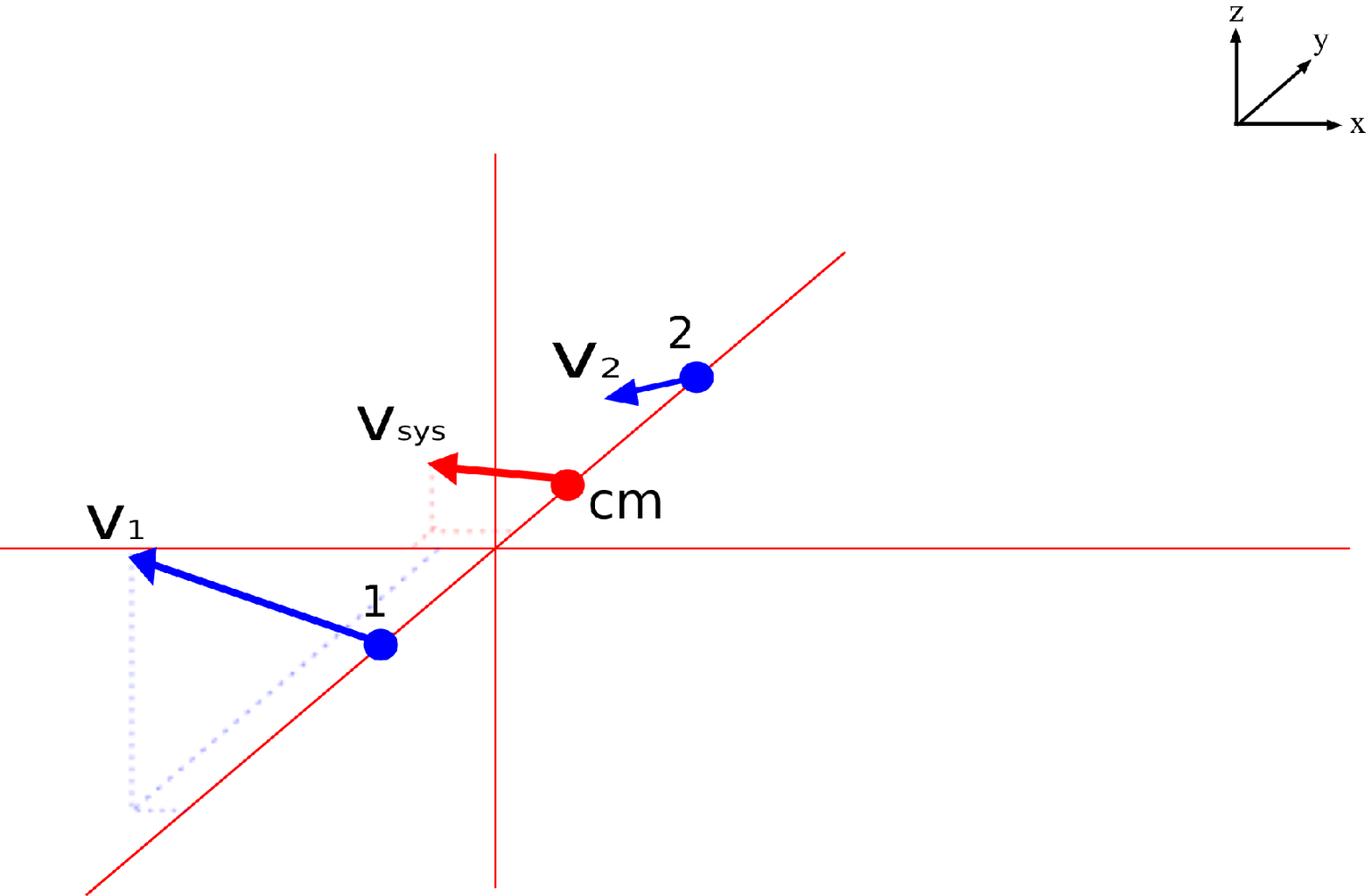}
  ~ b. The cm coordinate system in the cm reference frame for a binary
  system after the SN.
  \caption{Schematic representation of a binary system in the pre- and
    post-SN phase. The solid blue circles denote the primary and
    secondary star; the solid red cirle denotes the cm. The solid
    arrows denote the velocities the stars or cm have at that phase;
    the dashed arrows denote the velocity the SN imposes on the stars
    or cm which will change its velocity in the next phase. a. In the
    pre-SN phase the coordinate system is centered on the cm being at
    rest. b. In the post-SN phase the coordinate system is no longer
    centered on the cm - the cm has been translated in the
    y-direction, towards the secondary, and has gained a velocity $v_{sys}$. In
    both cases the inner binary orbital plane lies in the xy-plane and
    the y-axis is the line connecting the primary and the
    secondary. \label{fig:Binary}}
\end{figure}

After the SN the orbital parameters have changed to: semi-major axis,
$a$, eccentricity, $e$, and true anomaly, $\theta$. For a general
Kepler orbit of two objects with masses $m_1$ and $m_2$ respectively,
a relative velocity, $v$, semi-major axis, $a$, and separation
distance, $r$, the orbital energy conservation equation is
\begin{equation}\label{eq:v^2}
 v^2 = G(m_1 + m_2)\Bigl( \frac{2}{r} - \frac{1}{a}\Bigr),
\end{equation}
where $G$ is Newton's gravitational constant. The specific relative
angular momentum $\mathbf{h}$ is related to the orbital parameters as
follows

\begin{eqnarray}\label{eq:h^2}
|\mathbf{h}|^2 &=& |\mathbf{r} \times \mathbf{v}|^2 \\
               &=& G(m_1 + m_2)a(1 - e^2),
\end{eqnarray}
where the first equality holds for all Kepler orbits and the second
only applies to bound orbits. For thorough studies on SNe in a binary
system see \citet{1983ApJ...267..322H}, \citet{1996ApJ...471..352K},
and \citet{1998A&A...330.1047T}; the latter authors also take into
account the shell impact on the companion star using a method proposed
by \citet*{1975ApJ...200..145W}. Following the mentioned works as
guides for our calculations on the binary system we use a total pre-SN
mass of $M_0 = m_{1,0} + m_2$. Without loss of generality, we choose a
coordinate system in which at $t=0$ the orbit lies in the xy-plane,
the center of mass of the binary (cm) is at the origin, the y-axis is
the line connecting the primary and the secondary (the cm coordinate
system; see Figure~\ref{fig:Binary}), and we choose a reference frame
in which at $t=0$ the cm is at rest (the cm reference
frame). 

Before the SN the separation distance between the stars is
\begin{equation}\label{eq:r}
\mathbf{r} = \mathbf{r_1} - \mathbf{r_2} = \left(
             0,-\frac{a_0(1-e_0^2)}{1+e_0\cos\theta_0},0 \right).
\end{equation}
Using the following notation
\begin{eqnarray*}\label{eq:secondary_velocity_vector}
 x &=& a_0\sqrt{1-e_0^2}\cos\gamma_0\cos\theta_0 + a_0\sin\gamma_0\sin\theta_0,\\
 y &=& -a_0\sqrt{1-e_0^2}\cos\gamma_0\sin\theta_0 + a_0\sin\gamma_0\cos\theta_0,\\
 v_{0x} &=& v_0\frac{x}{\sqrt{x^2 + y^2}},\\
 v_{0y} &=& v_0\frac{y}{\sqrt{x^2 + y^2}},
\end{eqnarray*}
in which $\gamma_0$ is the pre-SN eccentric anomaly defined by $r =
a_0(1-e_0\cos\gamma_0)$, the velocity of the primary relative to the
secondary is
\begin{equation}\label{eq:v_0}
\mathbf{v_0} = \mathbf{v_{1,0}} - \mathbf{v_2} = (v_{0x},v_{0y},0).
\end{equation}
After the SN the primary has lost a part of its mass,
$\Delta$m, and has obtained a velocity kick $\mathbf{v_k}$ in a random
direction, which makes an angle $\phi$ with the pre-SN relative
velocity $\mathbf{v_0}$. The velocity of the primary relative to the
secondary, after the SN, is
\begin{equation}\label{eq:v}
\mathbf{v} = \mathbf{v_0} + \mathbf{v_k} = (v_{0x}+v_{kx},v_{0y}+v_{ky},v_{kz}), 
\end{equation}
the mass of the primary is $m_1 = m_{1,0} - \Delta m$ and the total
binary mass is $M = M_0 - \Delta m$. Applying these relations and
equations (\ref{eq:v^2}) and (\ref{eq:h^2}) to the binary system, we
obtain equations relating the post-SN semi-major axis, $a$, and
eccentricity, $e$, to both the pre- and post-SN orbital parameters and
velocities. Using $v_{c,0} = v_0|_{r=a_0} = (GM_0/a_0)^{1/2}$ as the
pre-SN relative velocity \citep{1983ApJ...267..322H}, we obtain
\begin{eqnarray}\label{eq:a}
 \frac{a}{a_0} &=& \Bigl(1-\frac{\Delta m}{M_0}\Bigr)\Bigl(1-\frac{2a_0}{r}\frac{\Delta m}{M_0}- 2\frac{v_0}{v_{c,0}}\frac{v_k}{v_{c,0}}\cos\phi \nonumber \\
 & & -\frac{v_k^2}{v_{c,0}^2}\Bigr)^{-1}
\end{eqnarray}
\begin{subequations}
\begin{eqnarray}
       e^2 &=& 1 - (1-e_0^2)
               \frac{M_0^2}{(M_0-\Delta m)^2}
                   \Bigl(1 -\frac{2a_0}{r}
                   \frac{\Delta m}{M_0} -\frac{v_k^2}{v_{c,0}^2} \nonumber \\ 
           & & - 2\frac{v_0}{v_{c,0}}\frac{v_k}{v_{c,0}}\cos\phi\Bigr)
               \label{eq:e^2}\\ 
           &=& 1 - \frac{a_0^2(1-e_0^2)^2}{a(1+e_0\cos\theta_0)^2}
                   \frac{(v_{0x}^2 +  v_{kx}^2 + v_{kz}^2 + 2v_{0x}v_{kx})}
                       {G(M_0 - \Delta m)}, \nonumber \\ 
           & & \label{eq:e^22}
\end{eqnarray}
\end{subequations}
which are consistent with \citet{1996ApJ...471..352K}.  In
\S\ref{Sect:ExampleTriple} we present a few examples regarding the
effect of mass loss and the supernova kick on the orbital parameters
of hierarchical triples.  To compute the systemic velocity of the
binary system due to the SN, we begin by writing the pre-SN velocities
of the primary and secondary in the cm reference frame; using the
pre-SN mass ratio $\mu_0 = m_2/M_0$, these velocities are given by
\begin{eqnarray}\label{eq:v_pre}
 \mathbf{v_{1,0}} &=& \mu_0 \Bigl(v_{0x}, v_{0y},0\Bigr)\label{eq:v_1_0},\\
 \mathbf{v_2} &=& (\mu_0-1) \Bigl(v_{0x},v_{0y},0\Bigr)\label{eq:v_2}.
\end{eqnarray}
As a result of the assumption of an instantaneous SN and neglecting
the shell impact, the instantaneous velocity of the secondary remains
unchanged after the SN, but the instantaneous velocity of the primary
changes to
\begin{eqnarray}\label{eq:v_post}
 \mathbf{v_1} = \Bigl(\mu_0 v_{0x} + v_{kx}, \mu_0 v_{0y} + v_{ky}, v_{kz}\Bigr)\label{eq:v_1}.
\end{eqnarray}
We now use the post-SN mass ratio $\mu = m_2/M$, and find the systemic velocity of the binary system:
\begin{eqnarray}\label{eq:v_sys}
 \mathbf{v_{sys}} &=& (1-\mu)\mathbf{v_1} + \mu \mathbf{v_2} \label{eq:10} \nonumber \\
                  &=& (1-\mu) \Bigl(\frac{\mu_0-\mu}{1- \mu}v_{0x} + v_{kx},\frac{\mu_0-\mu}{1- \mu}v_{0y} + v_{ky},v_{kz}\Bigr). \nonumber \\
                  & & 
\end{eqnarray}
These results are consistent with the previously mentioned studies
on SN in binaries. As a conseqence a binary in which the compact
object does not receive a kick in the supernova explosion moves
through space like a frisbee.

\subsubsection{Dissociating binary systems}\label{subsubsec:DissBinary}
The mass loss and the kick velocity have a potentially disrupting
effect on the binary system. However, in cases where the mass loss
alone would have been large enough to unbind the binary, the
combination of the two can result in the binary system surviving the
SN \citep{1983ApJ...267..322H}. If the binary system dissociates, the
two stars move away from each other on a hyperbolic or, in a limiting
case, a parabolic trajectory. This corresponds to the cases where $a <
0$ and $e > 1$ (hyperbola) or $a \to \infty$ and $e = 1$
(parabola). From equation~(\ref{eq:a}) we see that for a dissociating
binary the angle $\phi$ between the kick velocity $\mathbf{v_k}$ and
the pre-SN relative velocity $\mathbf{v_0}$ satisfies
\citep{1983ApJ...267..322H}:
\begin{eqnarray}\label{eq:dissociating_phi}
 \cos \phi &\geq& \Bigl(1 - \frac{2a_0}{r}\frac{\Delta m}{M_0} -
 \frac{v_k^2}{v_{c,0}^2}\Bigr)\Bigl(2\frac{v_k}{v_{c,0}}\sqrt{\frac{2a_0}{r}
   - 1}\Bigr)^{-1}.
\end{eqnarray}
If the right-hand side of equation (\ref{eq:dissociating_phi}) is less than $-1$,
the binary dissociates for all $\phi$; but if it is greater than $1$ the binary
survives for all $\phi$.  If the right-hand side is within the range
$-1$ to $1$, the probability of dissociating the binary is \citep{1983ApJ...267..322H}:
\begin{eqnarray}\label{eq:Pdiss}
 P_{diss} &=& \frac{1}{2}\Bigl(1 -\Bigl(1 - \frac{2a_0}{r}\frac{\Delta
   m}{M_0} - \frac{v_k^2}{v_{c,0}^2}\Bigr)\Bigl(2\frac{v_0}{v_{c,0}}\frac{v_k}{v_{c,0}}\Bigr)^{-1}\Bigr). \nonumber
 \\ & &
\end{eqnarray}
\citet{1998A&A...330.1047T} presented analytical formulas to calculate the
dissociation velocities for a binary with a pre-SN circular orbit. We follow
Tauris \& Takens' calculations, though ignore the SN shell impact,
to derive the runaway velocities of two stars in dissociating binaries,
however we do so for a pre-SN orbit with arbitrary eccentricity. We use the
cm coordinate system, explained above. Using the following shorthand relations
\begin{eqnarray*}
 \tilde{m} &=& \frac{M}{M_0}, \\
 j &=& \frac{v_{0x}^2}{v_0^2} - 2\tilde{m}\frac{a_0}{2a_0-r} + \frac{v_k^2}{v_0^2} + \frac{2v_{0x}v_{kx}}{v_0^2}, \\
 k &=& 1 + \frac{j}{\tilde{m}}\frac{2a_0-r}{a_0} - \frac{v_{ky}^2}{\tilde{m}v_0^2}\frac{2a_0-r}{a_0}, \\
 l &=& \frac{1}{\mu}\Bigl(\frac{\sqrt{j}}{\tilde{m}v_0}v_{ky}\frac{2a_0-r}{a_0}-\frac{j}{\tilde{m}}\frac{2a_0-r}{a_0} - 1\Bigr), \\
 n &=& \frac{1}{\mu}\Bigl(1+\frac{j}{\tilde{m}}\frac{2a_0-r}{a_0}(k+1)\Bigr),
\end{eqnarray*}
we find the runaway velocities for the primary and secondary star:
\begin{eqnarray}
 \mathbf{v_{1,diss}} &=& \Bigl(v_{kx}\Bigl(\frac{1}{l} + 1 \Bigr) + \Bigl(\frac{1}{l} + \mu_0\Bigr)v_{0x}, \mu_0 v_{0y}\nonumber \\ 
 & & + v_{ky}\Bigl(1-\frac{1}{n}\Bigr) + \frac{k\sqrt{j}}{n}v_0, v_{kz}\Bigl(\frac{1}{l} + 1\Bigr)\Bigr),\\
 \mathbf{v_{2,diss}} &=& \Bigl(-\frac{v_{kx}}{m_2l} - \Bigl(\frac{1}{m_2l} +1 -\mu_0\Bigr)v_{0x}, (\mu_0-1)v_{0y}\nonumber \\
 & & + \frac{v_{ky}}{m_2n} - \frac{k\sqrt{j}}{m_2n}v_0, -\frac{v_{kz}}{m_2l}\Bigr).
\end{eqnarray}
\subsection{Hierarchical triple systems}\label{subsec:Triple}
\begin{figure}
  \centering
  \includegraphics[width=\linewidth]{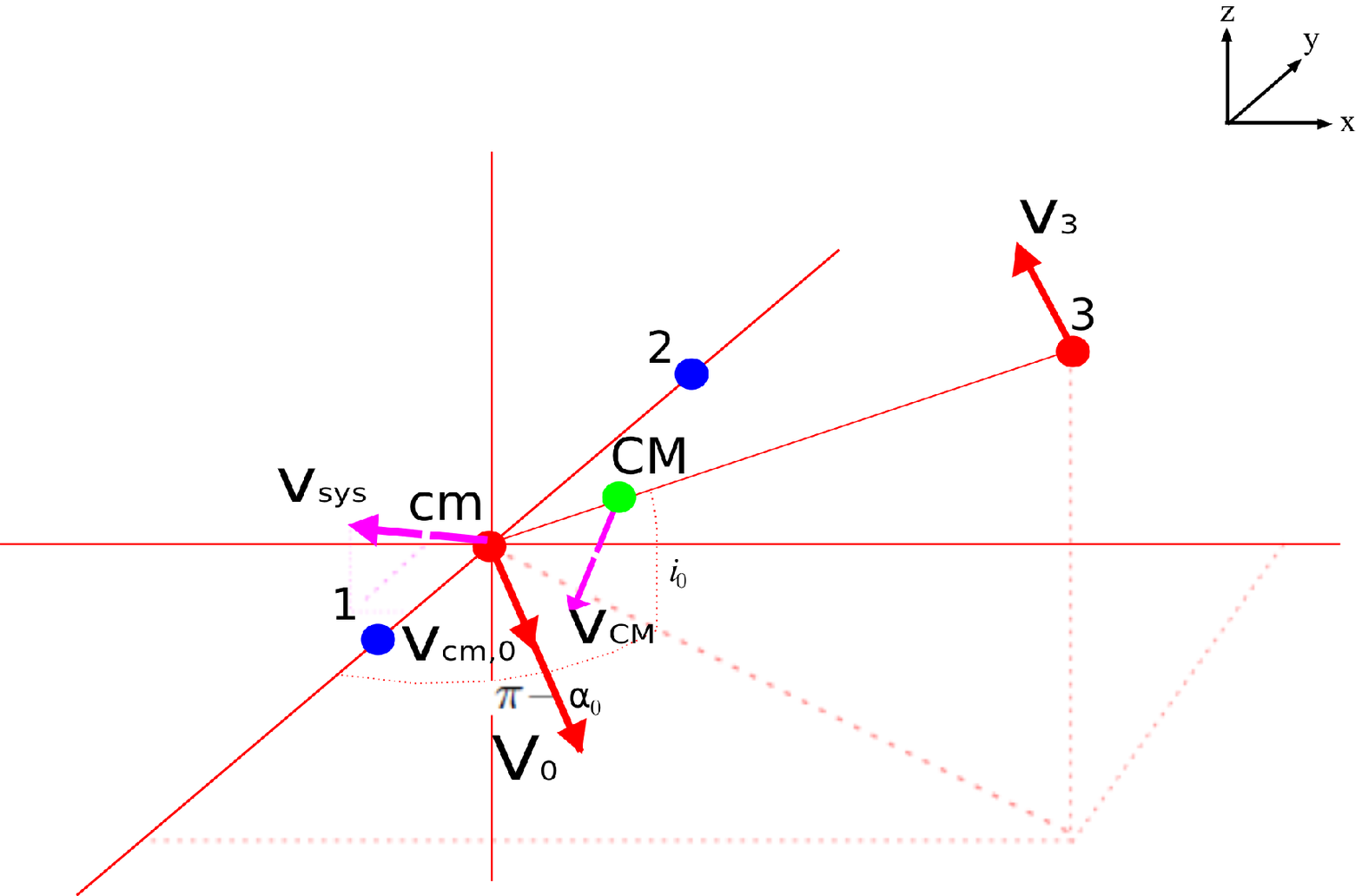}
  ~ a. The cm coordinate system in the CM reference frame for a
  hierarchical triple system before the SN (at $t = 0$).
  \includegraphics[width=\linewidth]{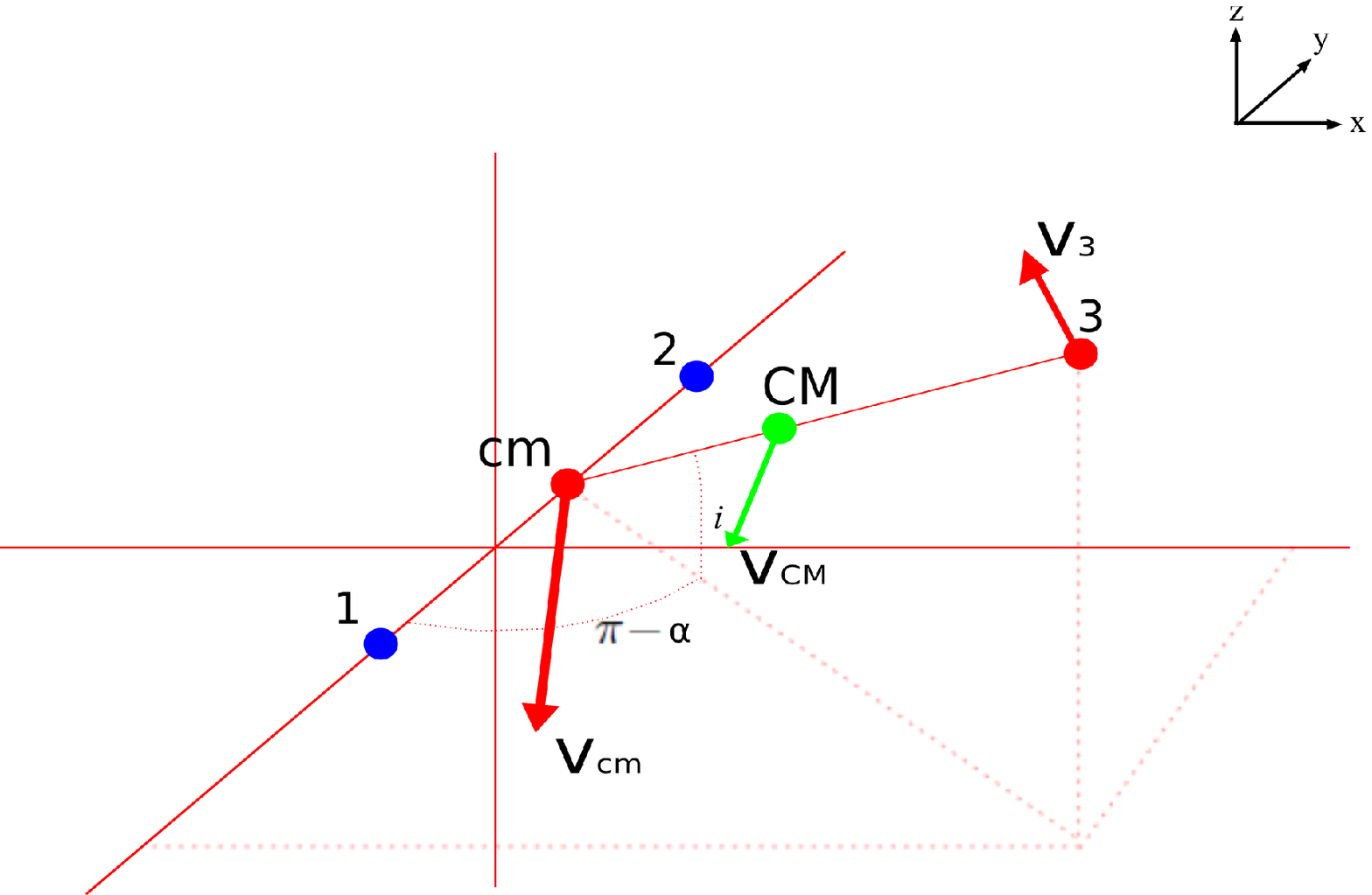}
  ~ b. The cm coordinate system in the CM reference frame for a
  hierarchical triple system after the SN.
  \caption{Schematic representation of a hierarchical triple star
    system in the pre- and post-SN phase. The solid blue circles
    denote the primary and secondary (inner binary); the solid red
    cirles denote the cm and the tertiary (outer binary); the green
    cirle denotes the CM. The solid arrows denote the velocities the
    stars or cm have at that phase; the dashed arrows denote the
    velocity the SN imposes on the stars or cm which will change its
    velocity in the next phase. (a) In the pre-SN phase at the moment
    immediately preceding the SN, the coordinate system is
    centered on cm and CM is at rest.  (b) In the post-SN phase the coordinate
    system is no longer centered on the cm - the cm has been translated in the
    y-direction, towards the secondary - and the CM is no longer at
    rest. In both cases the inner binary orbital plane lies in the
    xy-plane and the y-axis is the line connecting the primary and the
    secondary. \label{fig:Triple} }
\end{figure}

We now consider a hierarchical system of three stars with the primary,
secondary and tertiary star having mass, position and velocity given
by ($m_{1,0}$,$\mathbf{r_1}$,$\mathbf{v_{1,0}}$),
($m_2$,$\mathbf{r_2}$,$\mathbf{v_2}$) and
($m_3$,$\mathbf{r_3}$,$\mathbf{v_3}$) respectively. The primary star
undergoes a SN and the inner binary configuration and parameters are
the same as in section~\ref{subsec:Binary}.  The effective mass of the
inner binary's centre of mass (cm) is $m_{cm,0} = m_{1,0} + m_2 = M_0$,
and is at position
\begin{eqnarray}\label{eq:r_cm_0}
 \mathbf{r_{cm,0}} = (1-\mu_0)\mathbf{r_1} + \mu_0\mathbf{r_2}
\end{eqnarray}
and has a velocity
\begin{eqnarray}\label{eq:v_cm_0}
\mathbf{v_{cm,0}} = (1-\mu_0)\mathbf{v_{1,0}} + \mu_0\mathbf{v_2}.
\end{eqnarray}
The cm and tertiary constitute an outer binary defined by the
semi-major axis, $A_0$, eccentricity, $E_0$, and true anomaly,
$\Theta_0$. The separation distance between the cm and the tertiary
star we denote by $\mathbf{R_0}$. Before the SN the outer binary
orbital plane has an inclination $i_0$ with respect to the inner
binary and the separation distance of the outer binary projected onto
the xy-plane makes an angle $\alpha_0$ with the separation distance of
the inner binary. This inner-outer binary configuration is to some
extent acceptable, because the triple is hierarchical. This implies
that the separation distance of the cm and the tertiary is large
compared to the separation distance of the primary and secondary, i.e. $R_0 \gg r_0$,
so that the tertiary experiences gravitational influence of the inner
binary as if it was coming from one star at the cm. We assume an
instantaneous SN\footnote{See section~\ref{subsec:Binary} and note
  that the statements about the inner companion (the secondary) also
  hold for the outer companion (the tertiary).}. Due to the primary
undergoing a SN, the inner binary experiences a mass loss $\Delta m$
and an effective kick velocity is imparted to the cm: the systemic
velocity of the inner binary $\mathbf{v_{sys}}$ given by equation
\ref{eq:v_sys}. In addition, because of the reduction in mass of the
primary, the position of the cm has changed due to an instantaneous
translation along the y-axis
\begin{eqnarray}\label{eq:DeltaR}
 \mathbf{\Delta R} &=& \mathbf{r_{cm}} - \mathbf{r_{cm,0}} \nonumber \\
 &=& (\mu-\mu_0)\frac{a_0(1-e_0^2)}{1+e_0\cos\theta_0} \Bigl(0,1,0\Bigr).
\end{eqnarray}
The orbital parameters change as a result of the SN: the inner binary
parameters change according to the description in
section~\ref{subsec:Binary} and the outer binary orbital parameters
change to semi-major axis, $A$, eccentricity, $E$, and true anomaly,
$\Theta$. The hierarchical triple before the SN has a total mass
$M_{t,0} = M_0 + m_3$. We use the cm coordinate system to pin down the
inner binary and add to this coordinate system the tertiary at a position such that $R_0 \gg r_0$ 
(see Figure~\ref{fig:Triple}). We now select a reference frame in which the center of mass of 
the triple (CM) is at rest (the CM reference frame).

Prior to the SN the separation distance between the cm and the
tertiary is
\begin{eqnarray}\label{eq:R_0}
\mathbf{R_0} = \frac{A_0(1-E_0^2)}{1+E_0\cos\Theta_0}\Bigl(\cos i_0\sin\alpha_0,-\cos i_0\cos\alpha_0, \sin i_0\Bigr),
\end{eqnarray}
and, using the following shorthand notation
\begin{eqnarray*}
  X &=& A_0\sqrt{1-E_0^2}\cos\Gamma_0\cos\Theta_0 + A_0\sin\Gamma_0\sin\Theta_0\\
  Y &=& -A_0\sqrt{1-E_0^2}\cos\Gamma_0\sin\Theta_0 + A_0\sin\Gamma_0\cos\Theta_0\\
  X' &=& X\cos\alpha_0 - Y\cos i_0\sin\alpha_0\\
  Y' &=& X\sin\alpha_0 + Y\cos i_0\cos\alpha_0\\
  Z' &=& Y\sin i_0\\
  V_{0x} &=& V_0\frac{X'}{\sqrt{X'^2 + Y'^2 + Z'^2}}\\
  V_{0y} &=& V_0\frac{Y'}{\sqrt{X'^2 + Y'^2 + Z'^2}}\\
  V_{0z} &=& V_0\frac{Z'}{\sqrt{X'^2 + Y'^2 + Z'^2}}
\end{eqnarray*}
in which $\Gamma_0$ is the pre-SN outer orbit eccentric anomaly defined by $R_0 = A_0(1-E_0\cos\Gamma_0)$,
the velocity of the cm relative to the tertiary is

\begin{eqnarray}\label{eq:V_0}
\mathbf{V_0} = \mathbf{v_{cm,0}} - \mathbf{v_3} = (V_{0x},V_{0y},V_{0z}).
\end{eqnarray}
The effective kick velocity $\mathbf{v_{sys}}$ makes an angle $\Phi$
with the pre-SN relative velocity of the cm with respect to the
tertiary star $\mathbf{V_0}$. After the SN the separation distance
between the cm and the tertiary star is
\begin{eqnarray}\label{eq:R}
\mathbf{R} &=& \mathbf{R_0} + \mathbf{\Delta R}, \nonumber \\
&=& \frac{A_0(1-E_0^2)}{1+E_0\cos\Theta_0}\Bigl(\cos i_0\sin\alpha_0, (\mu-\mu_0)\frac{a_0(1-e_0^2)}{1+e_0\cos\theta_0}\nonumber \\
& & \times \frac{1+E_0\cos\Theta_0}{A_0(1-E_0^2)}-\cos i_0\cos\alpha_0,\sin i_0\Bigl),
\end{eqnarray}
the velocity of the cm relative to the tertiary star is
\begin{eqnarray}\label{eq:V}
\mathbf{V} &=& \mathbf{V_0} + \mathbf{v_{sys}} \nonumber \\
&=& (V_{0x} + v_{sys,x},V_{0y} + v_{sys,y},V_{0z} + v_{sys,z}),
\end{eqnarray}
the cm mass is $m_{cm} = M_0 - \Delta m$ and the total triple mass is
$M_t = M_{t,0} - \Delta m$. The inclination of the outer binary
orbital plane with respect to the inner binary orbital plane is given by:
\begin{eqnarray}\label{eq:i}
 \sin i = \frac{|\mathbf{R_0}|}{|\mathbf{R}|}\sin i_0.
\end{eqnarray}
The angle of the outer binary separation distance projected onto the xz-plane relative to
the inner binary separation distance is given by:
\begin{eqnarray}\label{eq:alpha}
 \sin\alpha = \frac{|\mathbf{R_0}|}{|\mathbf{R}|}\frac{\cos i_0}{\cos i}\sin\alpha_0.
\end{eqnarray}
Applying the relevant equations above and equations (\ref{eq:v^2}) and
(\ref{eq:h^2}) to our triple system, we obtain equations relating the
post-SN semi-major axis, $A$, and eccentricity, $E$, to both the pre- and
post-SN orbital parameters and velocities. Using $V_{c,0} =
V_0|_{R_0=A_0} = (GM_{t,0}/A_0)^{1/2}$ as the pre-SN relative velocity
when $R_0=A_0$, and using $\rho = (R_0-R)/(R_0 R)$, we obtain
\begin{eqnarray}\label{eq:A}
 \frac{A}{A_0} &=& \Bigl(1-\frac{\Delta m}{M_{t,0}}\Bigr)\Bigl(1-\frac{2A_0}{R}\frac{\Delta m}{M_{t,0}} - 2\frac{V_0}{V_{c,0}}\frac{v_{sys}}{V_{c,0}}\cos\Phi \nonumber \\
 & & -\frac{v_{sys}^2}{V_{c,0}^2} +2A_0\rho \Bigr)^{-1}, 
\end{eqnarray}
\begin{eqnarray}\label{eq:E}
 E^2 &=& 1 - (1-E_0^2)\frac{M_{t,0}}{(M_{t,0}-\Delta m)}\Bigl(\frac{2A_0}{R} + \frac{M_{t,0}}{M_{t,0}-\Delta m} \nonumber \\
 & & \times \Bigl(1 - \frac{2A_0}{R_0} -\frac{v_{sys}^2}{V_{c,0}^2} - 2\frac{V_0}{V_{c,0}}\frac{v_{sys}}{V_{c,0}}\cos\Phi\Bigr)\Bigr). \nonumber \\
 & &
\end{eqnarray} 
With the pre-SN mass ratio $\nu_0 = m_3/M_{t,0}$, the pre-SN velocities of the cm and the tertiary in the CM reference frame are
\begin{eqnarray}\label{eq:v_cm_0_in_terms_of_V0}
 \mathbf{v_{cm,0}} &=& \nu_0\Bigl(V_{0x},V_{0y},V_{0z}\Bigr)\\
 \mathbf{v_3} &=& (\nu_0 - 1)\Bigl(V_{0x},V_{0y},V_{0z}\Bigr).
\end{eqnarray}
We calculate the instantaneous velocity of the cm after the SN (as before, because of the assumption of an instantaneous SN, the velocity of the tertiary after the SN remains unchanged):
\begin{eqnarray}\label{eq:v_cm}
 \mathbf{v_{cm}} = \nu_0\Bigl(V_{0x} + \frac{v_{sys,x}}{\nu_0},V_{0y} + \frac{v_{sys,y}}{\nu_0},V_{0z} + \frac{v_{sys,z}}{\nu_0}\Bigr).
\end{eqnarray}
Using the post-SN mass ratio $\nu = m_3/M_t$, the systemic velocity of the outer binary (and therefore of the triple) is
\begin{eqnarray}\label{eq:V_sys}
 \mathbf{V_{sys}} &=& (1-\nu)\mathbf{v_{cm}} + \nu \mathbf{v_{3}} \nonumber \\
                  &=& (1-\nu)\Bigl(\frac{\nu_0 - \nu}{1-\nu}V_{0x} + (\mu_0-\mu)v_{0x} + (1-\mu)v_{kx}, \nonumber \\
                  & & \frac{\nu_0 - \nu}{1-\nu}V_{0y} + (\mu_0-\mu)v_{0y} + (1-\mu)v_{ky}, \nonumber \\
                  & & \frac{\nu_0 - \nu}{1-\nu}V_{0z} + (1- \mu)v_{kz}\Big).
\end{eqnarray}
Summarizing, one can consider a hierarchical triple system as a
effective binary system composed of an effective star (i.e. the inner
binary center of mass (cm)) and the tertiary. The effective star
undergoes an effective asymmetric SN resulting in three effects: 1)
sudden mass loss $\Delta m$, 2) an instantaneous translation
$\mathbf{\Delta R}$, and 3) a random kick velocity $\mathbf{v_{sys}}$.
The calculation of the post-SN parameters and velocities of a
hierarchical triple system is now reduced to the prescription for a SN
in a binary as presented in section~\ref{subsec:Binary}. Note that the
mass loss does not occur from the position of the effective star, but
from the position of the primary star; a clear distinction from a
physical binary system. However, from what position the mass loss
occurs is not important when an instantaneous SN is considered. When
the effect of the shell impact on the companion star(s) is considered,
this off-center mass loss must be taken into account. 
In addition, if it were not the primary which underwent the SN, but,
for example, the tertiary, the computation would have been done by
\textit{reducing} the inner binary to an effective star, as shown in this
section.  One would again have a binary configuration to calculate the effect
of the SN; in such a system there is no off-center mass loss. In
section~\ref{subsec:multiplicity} we show how one can reduce any
hierarchical multiple star system to an effective binary in a
recursive way using the effective binary method and in
\S\,~\ref{sec:AppendixBB} we do the computation of the effect of a SN
on a binary-binary system.

\subsubsection{Dissociating hierarchical triple systems}\label{subsubsec:DissTriple}
For the triple system, dissociation can occur in two ways: the inner
binary can dissociate ($a < 0$ and $e > 1$ or $a \to \infty$ and $e =
1$) (see section~\ref{subsec:Binary}) and the outer binary can
dissociate ($A < 0$ and $E > 1$ or $A \to \infty$ and $E = 1$),
i.e. the inner binary and the tertiary become unbound. The inner
binary dissociation scenario generally results in complete
dissociation of the system. However, hypothetical scenarios exist in
which one of the inner binary components is ejected towards the
tertiary star to either collapse with it or to form a binary by
gravitational or tidal capture. Nevertheless, these scenarios have a
small probability since the ejection conditions (e.g. the solid angle
in which that particular inner binary component has to be ejected in)
and the capture conditions are extremely specific. From
equation~\ref{eq:A} we see that for the inner binary to dissociate
from the tertiary, the angle $\Phi$ has to satisfy
\begin{eqnarray}\label{eq:dissociating_Phi}
\cos \Phi &\geq& \Bigl(1 - \frac{2A_0}{R}\frac{\Delta m}{M_{t,0}} - \frac{v_{sys}^2}{V_{c,0}^2}+ 2A_0\rho\Bigr)\Bigl(2\frac{V_0}{V_{c,0}}\frac{v_{sys}}{V_{c,0}}\Bigr)^{-1}.\nonumber\\
 & &
\end{eqnarray}
The probability of this type of dissociation is
\begin{eqnarray}\label{eq:Pdiss_outer}
 P_{diss}^{outer} &=& \frac{1}{2}\Bigl(1 -\Bigl(1 - \frac{2A_0}{R}\frac{\Delta m}{M_{t,0}} - \frac{v_{sys}^2}{V_{c,0}^2}+ 2A_0\rho\Bigr)\nonumber\\
 & & \times \Bigl(2\frac{V_0}{V_{c,0}}\frac{v_{sys}}{V_{c,0}}\Bigr)^{-1}\Bigr).
\end{eqnarray}
In the case of the dissociation of the outer binary, using the following short hand relations
\begin{eqnarray*}
 \tilde{M} &=& \frac{M_t}{M_{t,0}} \\
 J &=& \frac{V_{0x}^2}{V_0^2} - 2\tilde{M}\frac{A_0}{2A_0-R_0}\frac{R_0}{R} + \frac{v_{sys}^2}{V_0^2} + \frac{2V_{0x}v_{sys,x}}{V_0^2} \\
 K &=& 1 + \frac{J}{\tilde{M}}\frac{2A_0-R_0}{A_0}\frac{R}{R_0} - \frac{v_{sys,y}^2}{\tilde{M}V_0^2}\frac{2A_0-R}{A_0}\frac{R}{R_0} \\
 L &=& \frac{1}{\nu}\Bigl(\frac{\sqrt{J}}{\tilde{M}V_0}v_{sys,y}\frac{2A_0-R_0}{A_0}\frac{R}{R_0}\nonumber \\ & & -\frac{J}{\tilde{M}}\frac{2A_0-R_0}{A_0}\frac{R}{R_0} - 1\Bigr) \\
 N &=& \frac{1}{\nu}\Bigl(1+\frac{J}{\tilde{M}}\frac{2A_0-R_0}{A_0}\frac{R}{R_0}(K+1)\Bigr)
\end{eqnarray*}
the runaway velocities of the inner binary system and the tertiary are (following and generalizing \citet{1998A&A...330.1047T}): 
\begin{eqnarray}
\mathbf{v_{cm,diss}} &=& \Bigl(v_{sys,x}\Bigl(\frac{1}{L} + 1 \Bigr) + \Bigl(\frac{1}{L} + \nu_0\Bigr)V_{0x}, v_{sys,y}\Bigl(1-\frac{1}{N}\Bigr)\nonumber \\
 & & + \nu_0 V_{0y} + \frac{K\sqrt{J}}{N}V_0, v_{sys,z}\Bigl(\frac{1}{L} + 1\Bigr)\Bigr)\\
 \mathbf{v_{3,diss}} &=& \Bigl(-\frac{v_{sys,x}}{m_3L} - \Bigl(\frac{1}{m_3L} + 1 -\nu_0\Bigr)V_{0x}, (\nu_0-1)V_{0y}\nonumber \\
 & & + \frac{v_{sys,y}}{m_3N} - \frac{K\sqrt{J}}{m_3N}V_0, -\frac{v_{sys,z}}{m_3L}\Bigr).
\end{eqnarray}
Note that these equations are more general than the ones in
section~\ref{subsubsec:DissBinary}, because we cannot assume
$\mathbf{R}=\mathbf{R_0}$ in the triple case.

\subsection{An example of the effect of a supernova in a hierarchical triple}
\label{Sect:ExampleTriple}

For two simple sets of initial conditions we investigated the effect
of mass loss, $\Delta m$, and kick velocity, $\mathbf{v_k}$, on the
survivability of a triple system. We distinguish between four
different post-SN scenarios: (1) the triple survives as a whole ($e <
1$ and $E < 1$) with new orbital parameters, (2) the inner binary
survives and the third star escapes ($e < 1$ and $E > 1$), (3) the
inner binary dissociates and the outer binary survives ($e > 1$ and $E
< 1$) and (4) the triple completely dissociates ($e > 1$ and $E >
1$). The third scenario is a rather special case and can only be of
temporary nature: in this scenario, even though the inner binary has
just dissociated, the third star remains bound to the inner binary
center of mass. This is a temporal solution which eventually will lead
to the full dissociation of the triple, except in the extreme case in
which the tertiary star captures one of the ejected inner stars to
form a new binary system.

\begin{figure*}
\centering
\subfloat[e$_0$ = 0, E$_0$ = 0, $\theta_0$ = 0$^\circ$, $\Theta_0$ = 0$^\circ$, i$_0$ = 0, $\alpha_0$ = 0$^\circ$, \newline $\mathbf{\hat{v}_k}$ = (1,0,0)]
{\includegraphics[width=0.5\textwidth]{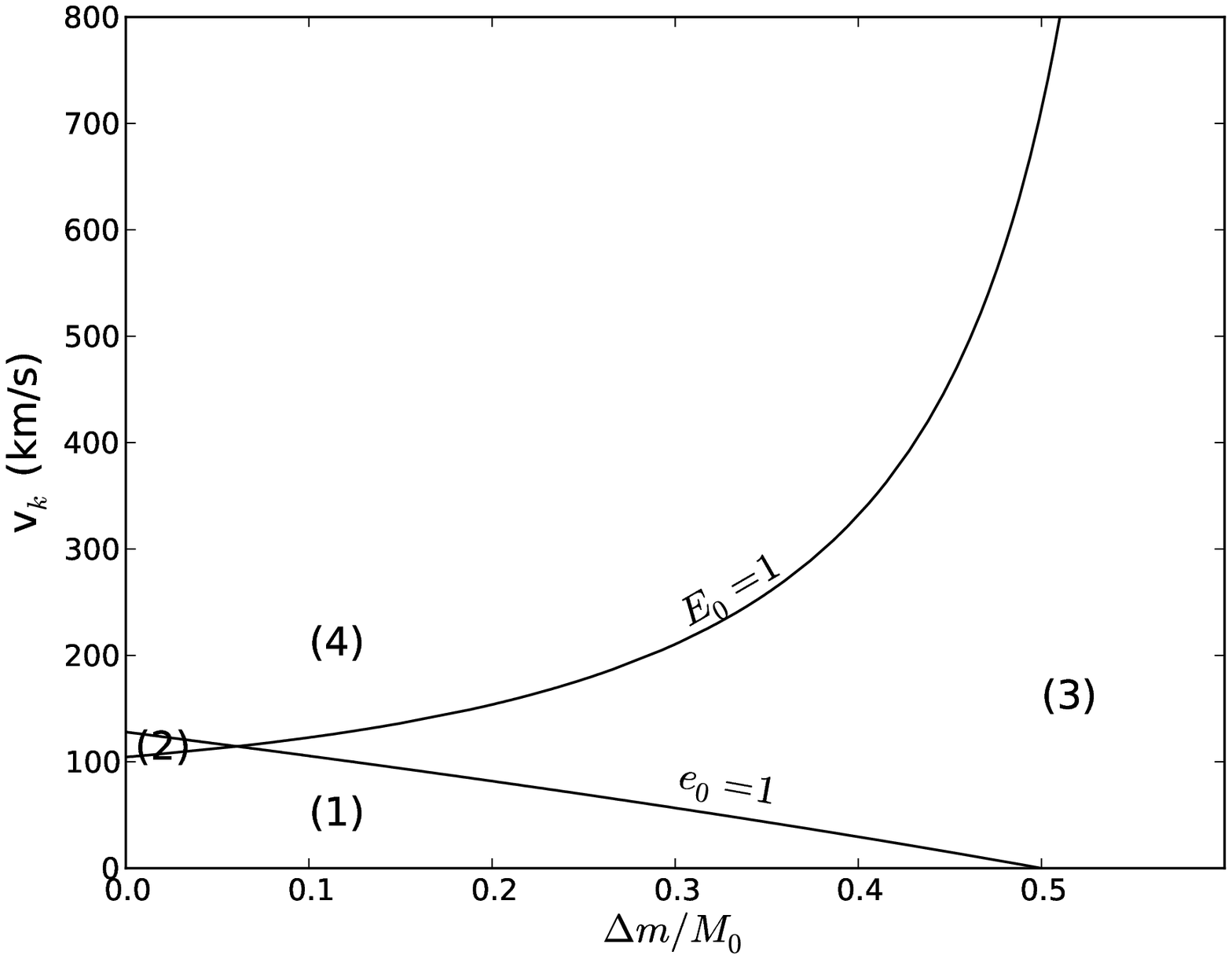} }
\centering
\subfloat[e$_0$ = 0, E$_0$ = 0, $\theta_0$ = 0$^\circ$, $\Theta_0$ = 0$^\circ$, i$_0$ = 0, $\alpha_0$ = 0$^\circ$, \newline $\mathbf{\hat{v}_k}$ = (-1,0,0)]
{\includegraphics[width=0.5\textwidth]{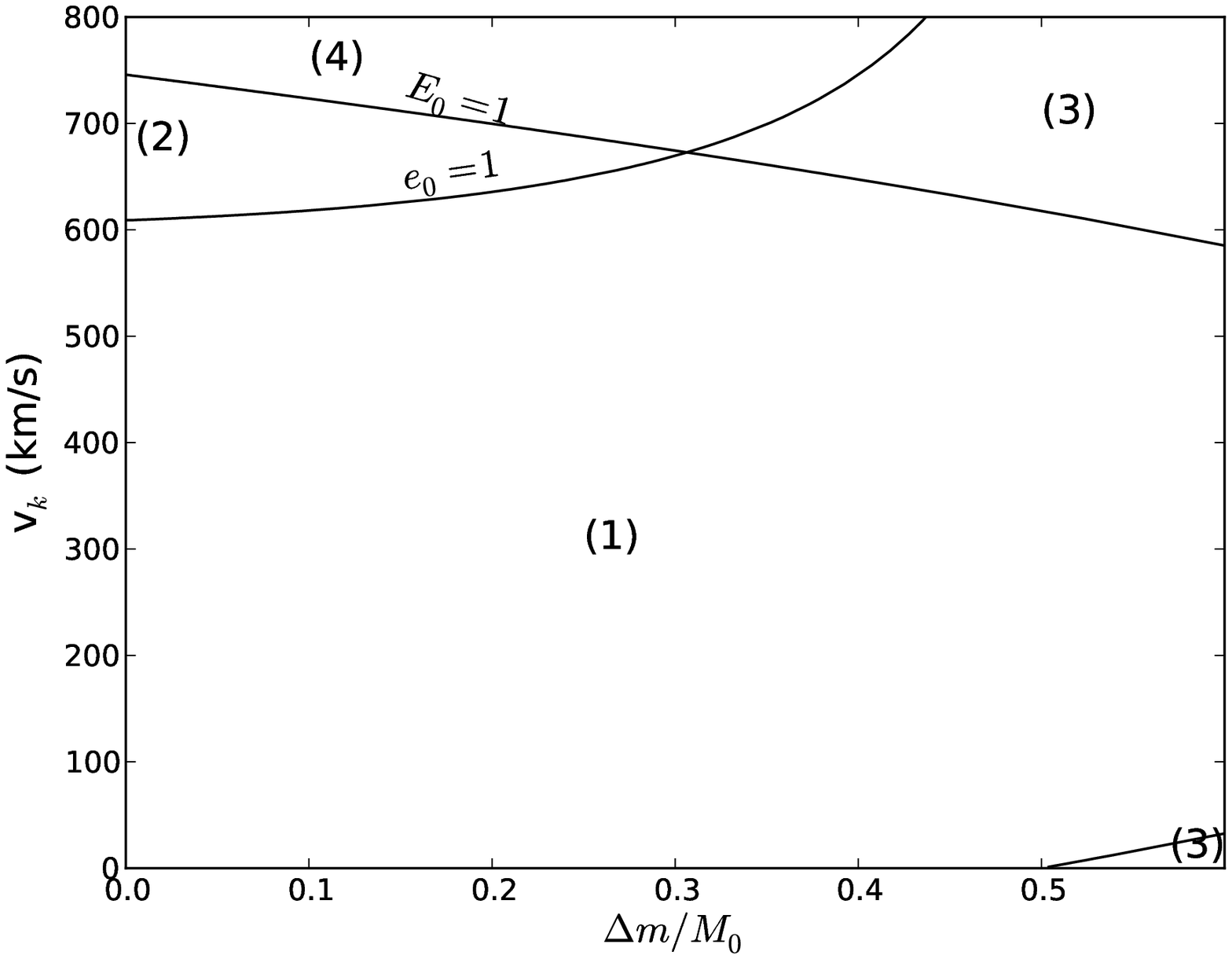} }
\caption{The plots above show the survivability of the hierarchical
  triple system for varying mass loss $\Delta m$ and kick velocity
  $v_k$.  The systems have masses of m$_{1,0}$, m$_2$, m$_3$ = 3, 2, 1
  M$_{\odot}$ respectively and inner and outer binary semi-major axes
  a$_0$, A$_0$ = 10, 50 R$_{\odot}$ respectively.  There are four
  possible post-SN scenarios: (1) the whole triple survives, (2) the
  inner binary survives but the third star escapes, (3) the inner
  binary dissociates and the outer binary survives, or (4) the triple
  completely dissociates.  The areas in the plots are labeled
  according to their respective post-SN
  scenario.\label{fig:v_k_vs_dM_over_M_0}}
\end{figure*}

For each set of initial conditions we used a hierarchical triple
system with primary, secondary and tertiary stars of masses m$_{1,0}$,
m$_2$, m$_3$ = 3, 2, 1 M$_{\odot}$ respectively and inner and outer
binary semi-major axes a$_0$, A$_0$ = 10, 50 R$_{\odot}$ respectively,
and we varied the kick velocity direction $\mathbf{\hat{v}_k}$. 
For the two different sets of initial conditions we determine which
combinations of $\Delta m$ and $v_k$ lead to which post-SN scenario and we
show our results in Figure \ref{fig:v_k_vs_dM_over_M_0}; the used initial
conditions are specified below the respective figures.

In Figure~\ref{fig:v_k_vs_dM_over_M_0}a. we used a circular inner and
outer orbit, not inclined with respect to each other, with all stars
on one line and the kick velocity in the same direction as the pre-SN
inner binary relative velocity. We see that for zero kick velocity,
the inner binary dissociates for a mass loss ratio of $\Delta m/M_0 =
0.5$, which is consistent with earlier work
(e.g. \citealt{1983ApJ...267..322H}). For zero mass loss, we see that
the inner binary dissociates for a kick velocity of $\mathbf{v_k} \sim
128$ km/s - this velocity is exactly the difference between the inner
binary escape velocity (v$_{esc} = \sqrt{2GM_0/a_0} \sim 437$ km/s)
and pre-SN relative velocity (v$_0 = \sqrt{GM_0/a_0} \sim 309$ km/s) -
but the third star escapes for a slightly lower value of the kick
velocity. This is because the inner binary systemic velocity (which is
the effective outer orbit kick; see Section~\ref{subsec:Triple}) plus
the pre-SN outer orbit relative velocity already exceed the outer
orbit escape velocity. We furthermore see that the total triple
survival scenario allows lower kick velocities for higher mass
losses. Above a kick velocity of $\mathbf{v_k} \sim 128$ km/s the
inner binary always dissociates, irrespective of the mass loss,
(eventually) leading to total dissociation.

In Figure~\ref{fig:v_k_vs_dM_over_M_0}b. we keep the same
configuration as described for Figure~\ref{fig:v_k_vs_dM_over_M_0}a.,
but with a kick velocity in the opposite direction with respect to the
orbital velocity of the exploding star before the supernova. The
triple can now lose more mass and receive a higher velocity kick while
stil surviving. The ability to sustain greater kick velocities is
explained by the fact that, depending on the mass loss, the kick
velocity now has to exceed a fraction of the sum of v$_0$ and v$_k$
(for zero mass loss v$_0$+v$_k$$\sim 746$ km/s) due to the opposing
directions of the two velocities. We also see that total triple
survival can occur beyond a mass loss ratio of 0.5, because the kick
velocity can oppose the dissociating effect of the mass loss (as
mentioned in \citealt{1983ApJ...267..322H}).  Bear in mind that while
the $\Delta m/M_0 = 0$ case is non-physical we include it for the sake
of completeness.

\begin{figure}
\subfloat{\includegraphics[width=0.5\textwidth]{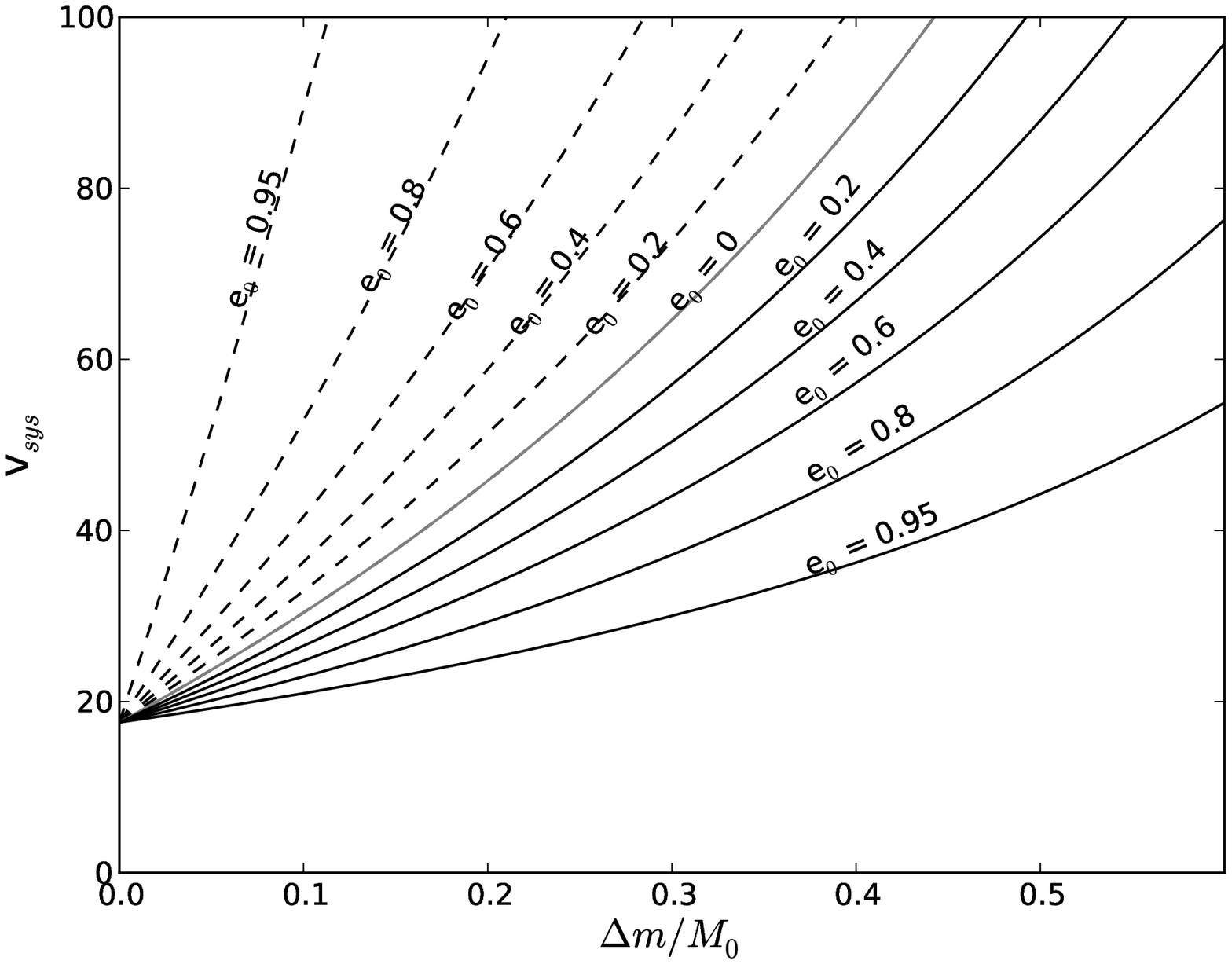} }\\
\subfloat{\includegraphics[width=0.5\textwidth]{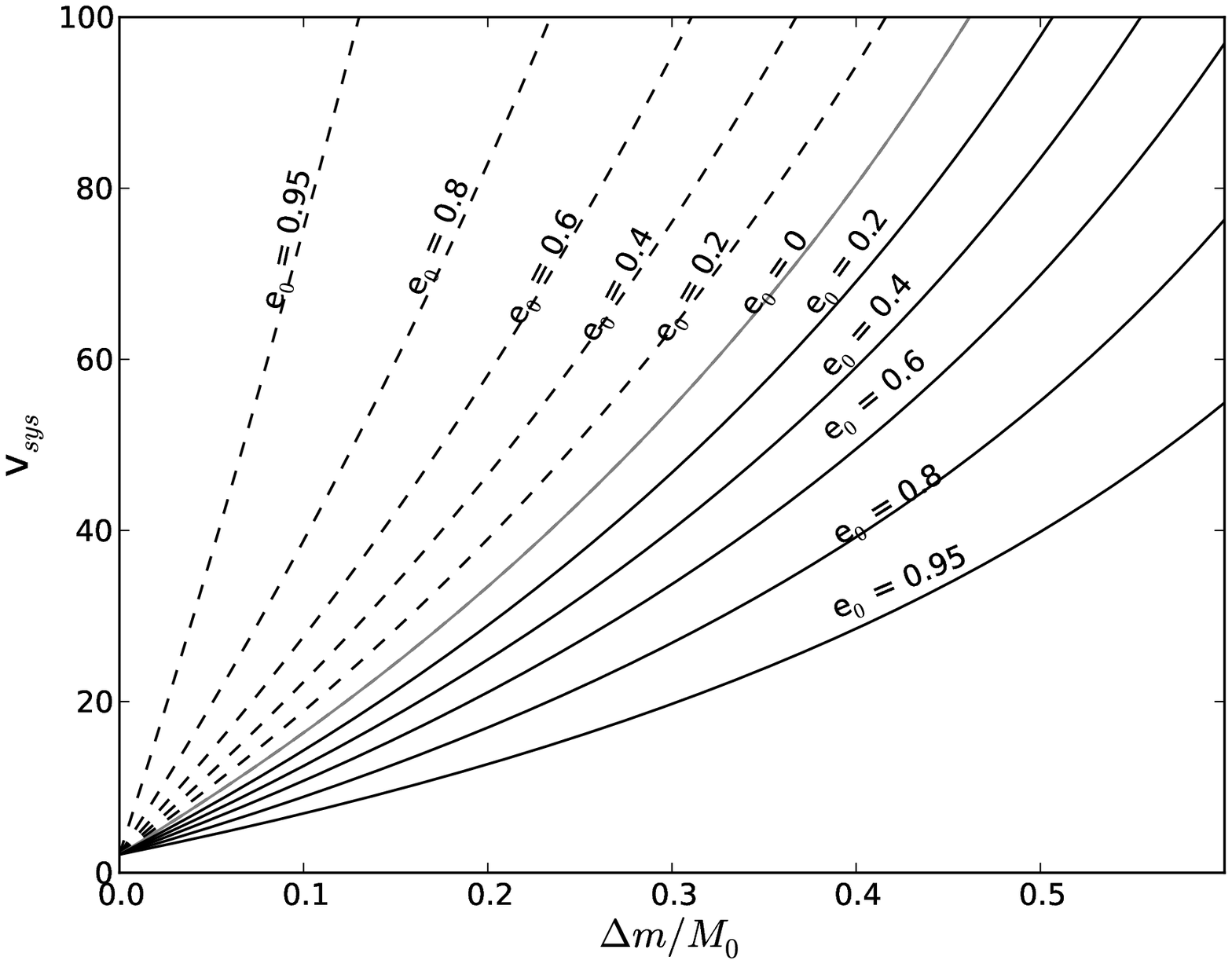} }
\caption{The post-SN systemic velocity of the triple as a function of mass
loss, $\Delta m$, when the SNe occurs at periastron ($\theta_0 = 0$, dashes)
and apastron ($\theta_0 = 180$ degrees, solid curves) of the inner binary,
for a range of pre-SN inner binary eccentricities.  $v_k = 0$ km/s in the top
panel and $\sim 31$ km/s in the bottom panel.\label{fig:vsys_vs_dM_over_M0}}
\end{figure}

In Figure~\ref{fig:vsys_vs_dM_over_M0} we show how the post-SN systemic
velocity of the triple depends on the mass loss $\Delta m$ for a
hierarchical triple system with primary, secondary and tertiary stars with
masses (m$_{1,0}$, m$_2$, m$_3$) = (3, 2, 1) M$_{\odot}$, inner and outer
binary semi-major axes (a$_0$, A$_0$) = (10, 50) R$_{\odot}$ and the kick
velocity in the direction of the pre-SN inner orbit relative velocity. We
plot our results for the case that the SN went off at the inner orbit
apastron ($\theta_0 = 180$ degrees) or at the inner orbit periastron
($\theta_0 = 0$ degrees) for a symmetric SN (i.e. $v_k = 0$ km/s) and a SN
with a kick $v_k \sim 31 $ km/s, in the cm reference frame (i.e. with the
cm at rest at $t = 0$).  In the top panel of Figure~\ref{fig:vsys_vs_dM_over_M0} we see
that for a symmetric supernova, the systemic velocity of the inner binary
increases with the amount of mass loss, which is an intuitive result. We
see that even with zero mass loss the triple has a systemic velocity,
namely the velocity it started with in this reference frame ($V_{sys} \sim
17.5$ km/s). We furthermore see that the increase of the triple systemic
velocity happens more steeply for these cases where the SN goes off at
periastron - with the steepest curve for the highest inner binary
eccentricity - than when the supernova goes off at apastron - with the
steepest curve is for lowest eccentricity. For an asymmetric supernova
with kick $v_k \sim 31 $ km/s (see the bottom panel of
Figure~\ref{fig:vsys_vs_dM_over_M0}) we observe similar behaviour, but with
the difference of the zero mass loss case: in this case the triple

system has a lower velocity than it started with ($V_{sys} \sim 2.5$
km/s), which is due to the kick. This result is dependent on the direction
of the kick.\\ The pre-SN triple systemic velocity is dependent on both
the inner binary and the outer binary. Its dependence on the inner binary
is via the masses $m_{1,0}$ and $m_2$ of the primary and secondary
respectively and the inner binary orbital parameters which fully constrain
the relative velocity of these stars (see equation (\ref{eq:v_0})). Its dependence on
the outer binary is via the mass $m_3$ of the tertiary and the outer orbit
orbital parameters which fully constrain the outer binary relative
velocity (see equation (\ref{eq:V_0})). The post-SN triple systemic velocity is
merely the sum of the pre-SN systemic velocity and its change, which is
only due to the inner binary through the mass loss $\Delta m$ and kick
velocity $\mathbf{v_k}$.

\subsection{Hierarchical systems of multiplicity $\bmath{ > 3}$}\label{subsec:multiplicity}
There exist two kind of hierarchical multiple star systems with more than three stars: 
\begin{enumerate}
 \item systems that have $n$ stars and hierarchy $n-1$, i.e. multiple star systems with its stars hierarchically ordered in series (hereafter serial systems). Examples of such systems include quadruples with hierarchy 3, but also binaries and triples are serial systems.
 \item systems that have $n$ stars and hierarchy $n-2$ or below, i.e. multiples composed of serial systems which are hierarchically ordered in parallel (hereafter parallel systems). An example of such system is a quadruple with hierarchy 2 (i.e. a binary-binary system).
\end{enumerate}
\subsubsection{Serial systems}\label{subsubsec:serial}
The effect of a SN on a serial system is calculated by applying the effective binary method (see section~\ref{subsec:Triple}) by recursively replacing the inner binary by an effective star at the center of mass of that binary, until the total system is reduced to a single effective binary. When considering a serial system of $n$ stars each with mass, position and velocity given by ($m_{1,0}$,$\mathbf{r_1}$,$\mathbf{v_{1,0}}$), ($m_2$,$\mathbf{r_2}$,$\mathbf{v_2}$), ... , ($m_n$,$\mathbf{r_n}$,$\mathbf{v_n}$) respectively, in which the primary star undergoes a SN, one starts by reducing the inner binary to an effective star, as was done in section~\ref{subsec:Triple}. The inner binary consists of the primary and secondary star at positions $\mathbf{r_1}$ and $\mathbf{r_2}$ respectively. This binary is reduced to an effective star of mass $m_{cm,0} = m_{1,0} + m_2$ at position $\mathbf{r_{cm,0}}$ given by equation (\ref{eq:r_cm_0}) and having velocity $\mathbf{v_{cm,0}}$ given by equation (\ref{eq:v_cm_0}). Due to the SN of the primary this effective star experiences a mass loss $\Delta m$, an instantaneous translation $\mathbf{\Delta R}$ given by equation (\ref{eq:DeltaR}), and a random kick velocity $\mathbf{v_{sys}}$ given by equation (\ref{eq:v_sys}). After applying these effects on this effective binary, one can calculate the post-SN orbital parameters and velocities and the systemic velocity $\mathbf{v_{sys}^{(2)}} = \mathbf{V_{sys}}$ of this effective binary, given by equation (\ref{eq:V_sys}), using the prescription for a SN in a binary.\footnote{The number between parentheses denotes the hierarchy up to which the system has been reduced to a effective star.} The total system is now reduced to a serial system of $n-1$ objects (real and effective stars).

Subsequently, one reduces the current inner binary - consisting of the effective and tertiary star at positions $\mathbf{r_{cm,0}}$ and $\mathbf{r_3}$ respectively - to an effective star of mass $m_{cm,0}^{(2)} = m_{cm,0} + m_3$, at position 
\begin{eqnarray}\label{eq:r_cm_0_2}
 \mathbf{r_{cm,0}^{(2)}} = \frac{m_{cm,0}\mathbf{r_{cm,0}} + m_3\mathbf{r_3}}{m_{cm,0}+m_3} 
\end{eqnarray}
with a velocity 
\begin{eqnarray}\label{eq:v_cm_0_2}
 \mathbf{v_{cm,0}^{(2)}} = \frac{m_{cm,0}\mathbf{v_{cm,0}} + m_3\mathbf{v_3}}{m_{cm,0}+m_3}.
\end{eqnarray}
Due to the SN of the primary star, this effective star also experiences a mass loss $\Delta m$, an instantaneous translation $\mathbf{\Delta R^{(2)}}$  - this time, the translation vector has non-zero y- and z-components - and a random kick velocity $\mathbf{v_{sys}^{(2)}}$. After applying these effects on this effective binary, one can calculate the post-SN orbital parameters and velocities and the systemic velocity $\mathbf{v_{sys}^{(3)}}$ of this effective binary using the prescription for a SN in a binary. The total system is now reduced to a serial system of $n-2$ objects (real and effective stars).

This procedure is carried on until the entire multiple is reduced to a single effective binary, consisting of the $n$th star at position $\mathbf{r_n}$ and a effective star of mass $m_{cm,0}^{(n-2)} = m_{cm,0}^{(n-3)} + m_{n-1}$ at position 
\begin{eqnarray}\label{eq:r_cm_0_n-2}
 \mathbf{r_{cm,0}^{(n-2)}} = \frac{m_{cm,0}^{(n-3)}\mathbf{r_{cm,0}^{(n-3)}} + m_{n-1}\mathbf{r_{n-1}}}{m_{cm,0}^{(n-3)}+m_{n-1}} 
\end{eqnarray}
with a velocity 
\begin{eqnarray}\label{eq:v_cm_0_n-2}
 \mathbf{v_{cm,0}^{(n-2)}} = \frac{m_{cm,0}^{(n-3)}\mathbf{v_{cm,0}^{(n-3)}} + m_{n-1}\mathbf{v_{n-1}}}{m_{cm,0}^{(n-3)}+m_{n-1}}.
\end{eqnarray}
This effective star also experiences mass loss $\Delta m$, an
instantaneous translation $\mathbf{\Delta R^{(n-2)}}$ and a random
kick velocity $\mathbf{v_{sys}^{(n-2)}}$. After applying these effects
on this (final) effective binary, one can calculate the post-SN
orbital parameters and velocities and the systemic velocity
$\mathbf{v_{sys}^{(n-1)}}$ for this effective binary (and therefore of
the total system) using the binary method.

When it is not the primary star which undergoes a SN, but the $m$th
star in the hierarchy, the procedure is carried out by first reducing
the inner serial system of $m-1$ stars to an effective star at its
center of mass. One can then apply the above explained method, as
there is no computational difference in whether the primary or the
secondary of a(n effective) binary undergoes the SN.
\subsubsection{Parallel systems}\label{subsubsec:parallel}
The effect of a SN on a parallel system is calculated by reducing each
parallel branch (which itself is a serial system) to an effective star
until an effective serial configuration is reached; after this, one
can use the method explained in the previous section. We consider a
parallel system of $i$ parallel branches, each consisting of an
arbitrary number $n_i$ of stars with mass, position and velocity
given by ($m_1$,$\mathbf{r_1}$,$\mathbf{v_1}$), ... ,
($m_{n_i}$,$\mathbf{r_{n_i}}$,$\mathbf{v_{n_i}}$) respectively, in
which the $m$th star - which is part of branch $j$ - undergoes a
SN. One starts by reducing all $i-1$ branches $\neq j$ to effective
stars. One then calculates the effect of the SN on branch $j$
(i.e. systemic velocity and mass loss) using the method described in
section~\ref{subsubsec:serial}. The total system is now reduced to an
effective serial system of $i$ effective stars in which the $j$th
effective star undergoes an effective SN with the systemic velocity of
branch $j$ as the kick velocity. The effect of this effective SN on
the total system, can be calculated by applying the method described
in section~\ref{subsubsec:serial} to this effective serial system. As
an example we will now demonstrate the effect of a SN on a
binary-binary system.

\subsubsection{An example of the effect of a supernova in binary-binary system}\label{sec:AppendixBB}

We consider a hierarchical binary-binary
system of stars with mass, position and velocity given by ($m_{1,0}$,$\mathbf{r_1}$,$\mathbf{v_{1,0}}$),
($m_2$,$\mathbf{r_2}$,$\mathbf{v_2}$),
($m_3$,$\mathbf{r_3}$,$\mathbf{v_3}$) and
($m_4$,$\mathbf{r_4}$,$\mathbf{v_4}$) respectively, in which the
primary star undergoes a SN. The binary consisting of the primary and
the secondary star (primary binary) has the configuration and the
parameters as in section~\ref{subsec:Binary} and has a center of mass
(cm$_1$, i.e. effective star 1) of mass $m_{cm1,0} = m_{1,0} + m_2 =
M_0$ at position given by equation (\ref{eq:r_cm_0}) with a velocity
$\mathbf{v_{cm1,0}}$ given by equation (\ref{eq:v_cm_0}). The
secondary binary consists of the tertiary and quaternary star and its
center of mass (cm$_2$, i.e. effective star 2) has a mass $m_{cm2} =
m_3 + m_4 = M_2$, is at position
\begin{eqnarray}\label{eq:r_cm2}
 \mathbf{r_{cm2}} = (1-\kappa)\mathbf{r_3} + \kappa\mathbf{r_4} \nonumber
\end{eqnarray}
and has velocity 
\begin{eqnarray}\label{eq:v_cm2}
\mathbf{v_{cm2}} = (1-\kappa)\mathbf{v_3} + \kappa\mathbf{v_4}, \nonumber
\end{eqnarray}
before the SN, where $\kappa = \frac{m_4}{M_2}$. The cm$_1$ and cm$_2$
constitute an effective binary defined by semi-major axis, $A_0$,
eccentricity, $E_0$, and true anomaly, $\Theta_0$. The separation
distance is denoted by ${\mathbf R}_0$. Before the SN the effective
binary orbital plane has inclination $i_0$ with respect to the primary
binary orbital plane and the separation distance of the effective
binary projected onto the xy-plane makes an angle $\alpha_0$ with the
separation distance of the primary binary. We assume an instantaneous
SN\footnote{See section~\ref{subsec:Binary} and note that these
  statements about the inner companion (secondary) star also hold for
  the outer companion (tertiary and quaternary) stars.}. In the
effective SN the cm$_1$ experiences a mass loss $\Delta m$, an
instantaneous translation $\mathbf{\Delta R}$ along the x-axis given
by equation (\ref{eq:DeltaR}) and a random kick velocity
$\mathbf{v_{sys}}$ given by equation (\ref{eq:v_sys}). The orbital
parameters change as a result of the SN: the primary binary parameters
change according to the description in section~\ref{subsec:Binary} and
the effective binary orbital parameters change to semi-major axis $A$,
eccentricity $E$ and true anomaly $\Theta$; the secondary binary
orbital parameters do not change when SN-shell impact is not taken
into account. Before the SN the binary-binary system has a total mass
$M_{bb,0} = m_{cm1,0} + m_{cm2}$, we use the cm$_1$ coordinate system
to pin down the primary binary and add to this coordinate system the
tertiary and quaternary at a position such that $R_0 \gg r_0$, and we choose a
reference frame in which the center of mass of the total binary-binary
system (CM$_{bb}$) is at rest (the CM$_{bb}$ reference frame) and in
which the cm$_1$ is at the origin at $t=0$. The separation distance
between the cm$_1$ and the cm$_2$, $\mathbf{R_0}$, is given by
equation (\ref{eq:R_0}) and the velocity of the cm$_1$ relative to the
cm$_2$ is
\begin{eqnarray}\label{eq:V_0_bb}
\mathbf{V_0} = \mathbf{v_{cm1,0}} - \mathbf{v_{cm2}} = (V_{0x},V_{0y},V_{0z}) 
\end{eqnarray}
prior to the SN. The effective kick velocity $\mathbf{v_{sys}}$ makes
an angle $\Phi$ with the pre-SN relative velocity
$\mathbf{V_0}$. After the SN the separation distance between the
cm$_1$ and the cm$_2$ is $\mathbf{R}$ given by equation (\ref{eq:R})
and the velocity of the cm$_1$ relative to the cm$_2$ is $\mathbf{V}$
given by equation (\ref{eq:V}), the cm$_1$ mass $m_{cm1} = m_{cm1,0} -
\Delta m = M$ and total binary-binary mass $M_{bb} = m_{cm1} + m_{cm2}
= M + M_2$. Applying the relations above and equations (\ref{eq:v^2})
and (\ref{eq:h^2}) to our binary-binary system, we obtain relations
for the post-SN semi-major axis $A$ and eccentricity $E$ in terms of
both the pre- and post-SN orbital parameters and velocities given by
equations (\ref{eq:A}) and (\ref{eq:E}) respectively with $M_{t,0}$
replaced by $M_{bb,0}$. To compute the systemic velocity due to the
SN, we express the pre-SN velocities of the cm$_1$ and the cm$_2$ in the
CM$_{bb}$ reference frame. Using the pre-SN mass ratio $\lambda_0 =
\frac{m_{cm2}}{M_{bb,0}}$, the pre-SN velocities are given by
\begin{eqnarray}\label{eq:v_cm_2_bb}
 \mathbf{v_{cm1,0}} &=& \lambda_0\Bigl(V_{0x},V_{0y},V_{0z}\Bigr)\\
 \mathbf{v_{cm2}} &=& (\lambda_0 - 1)\Bigl(V_{0x},V_{0y},V_{0z}\Bigr).
\end{eqnarray}
We calculate the instantaneous velocity of the cm$_1$ after the SN (due to the assumption of an instantaneous SN, the velocity of the cm$_2$ after the SN remains unchanged):
\begin{eqnarray}
 \mathbf{v_{cm1}} = \lambda_0\Bigl(V_{0x} + \frac{v_{sys,x}}{\lambda_0},V_{0y} + \frac{v_{sys,y}}{\lambda_0},V_{0z} + \frac{v_{sys,z}}{\lambda_0}\Bigr)
\end{eqnarray}
With the post-SN mass ratio $\lambda = \frac{m_{cm2}}{M_{bb}}$, the systemic velocity of the effective binary (and therefore of the binary-binary system) is
\begin{eqnarray}\label{eq:V_sys_bb}
 \mathbf{V_{sys}} &=& (1-\lambda)\mathbf{v_{cm1}} + \lambda \mathbf{v_{cm2}} \nonumber \\
                  &=& (1-\lambda)\Bigl(\frac{\lambda_0 - \lambda}{1-\lambda}V_{0x} + (\mu_0-\mu)v_{0x} + (1-\mu)v_{kx}, \nonumber \\
                  & & \frac{\lambda_0 - \lambda}{1-\lambda}V_{0y} + (\mu_0-\mu)v_{0y} + (1-\mu)v_{ky}, \nonumber \\
                  & & \frac{\lambda_0 - \lambda}{1-\lambda}V_{0z} + (1- \mu)v_{kz}\Big).
\end{eqnarray}
Note that because the branch harboring the SN-progenitor (SN branch) is a binary, this calculation the SN-effect on the binary-binary system is almost identical to calculation of the SN-effect on a hierarchical triple. The computations become more interesting for systems with a SN branch of higher multiplicity.

\section{Application: Formation of J1903+0327}\label{sec:J1903}
PSR J1903+0327 was observed by \citet{2008Sci...320.1309C} who
determined it to be a millisecond pulsar (MSP). This MSP is observed
to have a 1 $M_\odot$ main sequence companion with a highly eccentric
and distant orbit ($e$ $\simeq$ 0.44, orbital period
$\simeq$ 95.2 days). These properties are atypical for MSPs because
MSPs are expected to be spun-up via mass transfer
\citep{1991PhR...203....1B}, which in turn widens and circularizes the
orbit, while its companion evolves through a giant
phase. \citet{1992RSPTA.341...39P}, for example, suggest an
eccentricity $e < 10^{-3}$ is typical for MSP binaries. The exception
to this has been MSPs in globular clusters which have interactions
with other objects that may perturb the orbit of the binary. However,
\citet{2011MNRAS.412.2763F} find it to be unlikely that this MSP
system has its origin in an exchange interaction in such a dense
stellar environment.

It has been suggested that J1903+0327 maybe the result of a
hierarchical triple (\citealt{2008Sci...320.1309C},
\citealt{2011ApJ...734...55P} and \citealt{2011A&A...536A..87B}) where
the inner companion has been lost after spinning-up the MSP, leaving
only the MSP and the former tertiary to be observed. Should J1903+0327
be the result of such a system the methods in the previous sections
provide a strong beginning to investigate how such a system might
evolve.

\subsection{Initial conditions}

We generate sets of $10^5$ initial conditions, as described below,
with each set constituting a stable triple system, and then simulated
the effect of an instantaneous SN occurring at the primary star. The
model we follow (many of our initial conditions are drawn from
  \citet{2011ApJ...734...55P}) consist of a primary, secondary and tertiary star with
zero age masses of 10 $M_\odot$, 1 $M_\odot$ and 0.9 $M_\odot$
respectively. The initial conditions are generated by selecting the
semi-major axis, $A_0$, eccentricity, $E_0$, and the orbital inclination,
$i$, for the tertiary.  $A_0$ takes values on the range
[200, 10\,000]$R_\odot$ from a flat distribution, $E_0$ is chosen on the
range [0, 1) from a distribution that is flat in log space, and $i_0$ is
  chosen on the range [0, $\pi$] with a sinusoidal distribution.
  Combining these values with the zero age masses of the stars as well
  as a pre-set value for the initial semi-major axis of the inner
  binary, $a_0 = 200R_\odot$ we then test for stability of the
  system using:
\begin{eqnarray}\label{eq:stability}
 \frac{A_0(1-E_0)}{a_0} &>& 3\Bigl(1+\frac{m_3}{M_0}\Bigr)^{1/3} \Bigl(\frac{7}{4}+\frac{1}{2}\cos i_0-\cos^2 i_0\Bigr)^{1/3} \nonumber \\
 &\times& (1-E_0)^{-1/6}
\end{eqnarray}
 \citep{2010ARep...54...38Z}.  If the system is stable with this set
 of parameters, we choose the remaining parameters, namely the angle
 $\alpha_0$ described in the previous sections, the direction and
 magnitude of the kick.  Because we have assured that the system is
 dynamically stable before starting our simulations our assumption of
 a hierarchical system is guaranteed.  We observe that due to the SN
 kick, systems with very high inclination are preferentially removed or 
 their inclination is reduced thus as a result we do not include the effects 
 of Kozai iterations.
\subsection{Simulations}
The inner binary undergoes a common envelope (CE) phase, circularizing
the orbit, reducing the inner semi-major axis to a value between 5 $R_\odot$
and 60 $R_\odot$, and reducing the mass of the primary to 2.7
$M_\odot$.  The effect of these changes on the stability of the system can immediately be seen in 
equation (\ref{eq:stability}).  Then, due to the SN, the primary undergoes a mass loss of
1.3 $M_\odot$ and receives a corresponding kick.  The velocity of the
kick is fixed between 5 and 160 km/s for each set of simulations and
the kick direction is randomly chosen such that for all simulations
the direction is isotropic. We then analyze the survivability and
stability of each system. A system survives the SN and resulting kick
if it remains bound, and it is determined to be stable if, while
remaining bound, the system also satisfies the stability criterion in
equation (\ref{eq:stability}).

\begin{figure}
  \centering
 \includegraphics[angle=270, width=\linewidth]{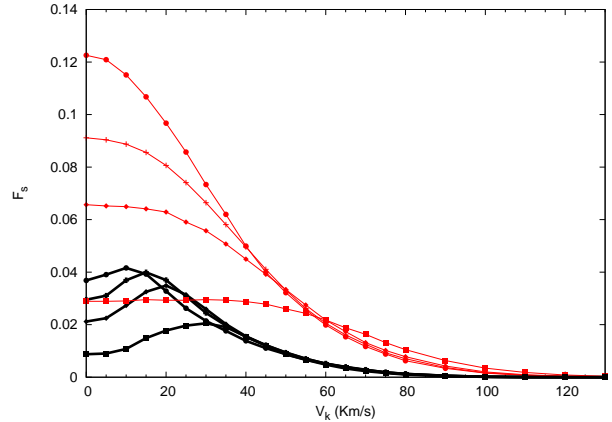}
  \caption{The fraction of surviving and stable system (thin red and
    thick black (colors online only) lines respectively) as a function
    of the kick velocity. The lines in each set correspond to
    different semi-major axis, 50, 30, 20, and 10 $R_\odot$ (circle,
    cross, diamond, and square respectively).  All curves are
    normalized to the total number of surviving systems with a
    semi-major axis of 50 $R_\odot$.\label{fig:sim1}}
\end{figure}
We ran Monte Carlo simulations for four different inner binary
semi-major axes (10, 20, 30, and 50 $R_\odot$).  For each semi-major
axis value we run 25 simulations (each of the 25 simulations consists
of $10^5$ sets of initial conditions) each with a constraint kick
velocity (between 0 and 130 km/s).  In Figure~\ref{fig:sim1} we plot
the kick velocity versus the fraction of surviving and stable
systems. For each pair of curves the thin red upper curve corresponds
to the survivability fraction and the thick black lower curve to the
fraction that survives and remains stable.  Curves with same kick
velocity have the same point-symbols. Each point represents the
fraction of surviving or stable systems normalized to the total number
of surviving systems with a semi-major axis of 50$R_\odot$. Increasing
the semi-major axis from 10 to 30 $R_\odot$ strongly increases the
overall probability of a system to survive and remain stable. However,
with a kick velocity of 45 km/s and higher the probability of a system
remaining stable is nearly the same when the semi-major axis is
$\geq$ 20$R_\odot$.  Figure~\ref{fig:sim1} shows the effect of the
Blaauw $\&$ Boersma recoil (\citealt{1961BAN....15..265B} $\&$
\cite{1961BAN....15..291B}) on the system when the SN kick is small;
as the SN kick velocity approaches the Blaauw $\&$ Boersma recoil
velocity the stability increases due to the kick and recoil
off-setting one another, in part or in full.  As the SN kick velocity
increases it begins to overwhelm the Blaauw $\&$ Boersma effect.

\begin{figure}
  \centering
 \includegraphics[angle=270, width=\linewidth]{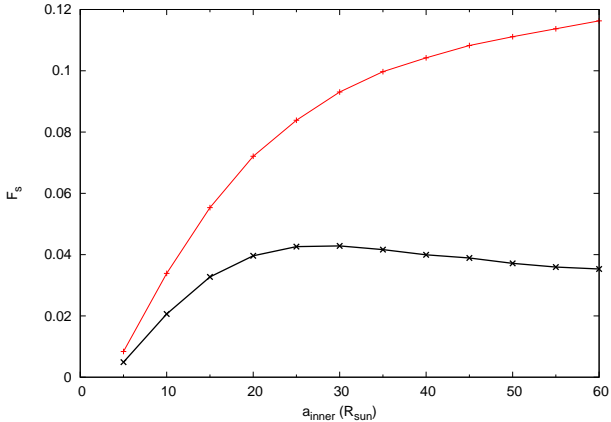}
  \caption{The fraction of surviving and stable systems (upper and lower lines respectively) with respect to the inner semi-major axis.  A constant kick velocity of 20 km/s is used.\label{fig:sim2}}
\end{figure}

In Figure~\ref{fig:sim2} we show the effect the inner semi-major axis
has on survivability and stability (the upper and lower lines
respectively) using a constant kick velocity of 20 km/s. Again each
data point represents the fraction of systems that survive or survive
and in addition remains stable out of a set of $10^5$ initial
conditions. Here we see the significant role of the inner semi-major
axis on the survivability of the system.  If we note for a particular
kick velocity which value of $a_0$ the stability fraction begins to
level, we can see it corresponds to the merging of the stability
curves in Figure~\ref{fig:sim1}.  For the case of a 20 km/s SN kick
velocity, as in Figure~\ref{fig:sim2}, we see that any value of $a_0$
greater than about 30 $R_\odot$ will have similar stability fractions
while systems with lower values of $a_0$ should have a lower stability
fraction as we see in Figure~\ref{fig:sim1}.

\begin{figure}
  \centering
 \includegraphics[angle=270, width=\linewidth]{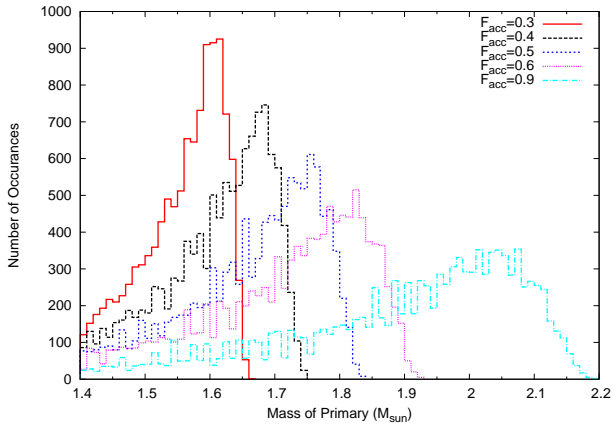}
  \caption{The number of occurrences for which the system becomes unstable due to mass transfer at a given mass of the primary.  The curves corresponds to $F_{acc}$ values of 0.3, 0.4, 0.5, 0.6, and 0.9 as shown in the key.  The peak value and FWHM for each curve in this figure, as well as similar curves for other values of $F_{acc}$, are plotted in Figure~\ref{fig:AccRate}.\label{fig:sim3}}
\end{figure}
Next, we chose all of the systems that remain stable after the SN and subject them to a mass transfer phase.  Here we iteratively remove one one-hundredth of the mass of the secondary and transfer a fraction of it to the primary, which after the SN would have formed a neutron star (NS).  Following the work of \citet{1994A&A...288..475P} we find: 
\begin{equation}\label{eq:aFinal}
 a_f = a_i\Bigl[\Bigl(\frac{m_{1,f}}{m_{1_i}}\Bigr)^{(1/(1-\chi))}\frac{m_{2,f}}{m_{2_i}}\Bigr]^{-2}\times\Bigl(\frac{M_{i}}{M_{f}}\Bigr)
\end{equation}
where $a_f$ is the new semi-major axis, $a_i$ is the semi-major axis before the mass transfer, $m_{1,i}$ and $m_{2,i}$ are the masses of the primary and secondary before the mass transfer and $m_{1,f}$ and $m_{2,f}$ are the masses of the primary and secondary after the mass transfer, $M_{i}$ and $M_{f}$ are the total masses of the binary before and after the mass transfer, and finally $\chi$ is the ratio of the change in mass of the system to the change in mass of the donor (i.e. the secondary).  If we define the fraction of mass accreted, $F_{acc}$, as the fraction of mass lost from the secondary which is accreted onto the primary we find that the $1/(1-\chi)$ term simply becomes $1/F_{acc}$.  
\begin{figure}
 \includegraphics[angle=270, width=\linewidth]{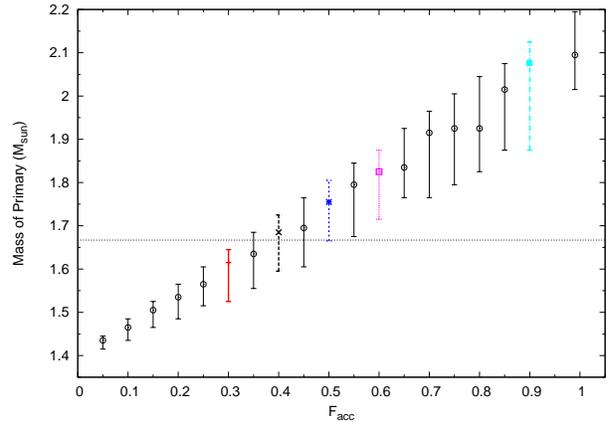}
  \caption{The final mass of primaries with respect to the fraction of accreted mass.  The dashed horizontal line is placed at the observed mass of J1903+0327.  The points represent the peak value of curves which plot the number of times a system becomes unstable while at a given mass of the primary (like those in Figure~\ref{fig:sim3}); the upper and lower bars represent the FWHM of the curves.  The values that are colored (online) and that have different line types correspond to the curves in Figure~\ref{fig:sim3} (e.g. the blue, dot-dash line at $F_{acc}$=0.9 is obtained from the right most peaked curve in Figure~\ref{fig:sim3}, which is also a blue, dot-dash line).\label{fig:AccRate} }
\end{figure}
After each iterative mass transfer, and the resulting change in the
semi-major axis, we test the triple for stability using equation
(\ref{eq:stability}).  When the system becomes dynamically unstable we
stop simulating as the assumption of a hierarchical system has broken
down.  We record the mass of the primary when the system becomes
dynamically unstable and plot the mass in Figure~\ref{fig:sim3}
versus the number of times systems becomes unstable at that mass.  For
this plot we used $F_{acc}$ values of 0.3, 0.4, 0.5, 0.6 and 0.9,
which correspond to the lines which peak from the left to right
respectively, and a constant kick velocity. We see that the peak value
for each $F_{acc}$ shifts to a larger primary mass as $F_{acc}$
increases.  This relation is expected since as $F_{acc}$ becomes
larger more of the mass lost from the secondary is accreted onto the
primary.  So for the case of $F_{acc} = 0.3$ only $30\%$ of the mass
lost from the secondary could ever accrete onto the primary thereby
reducing the maximum possible mass of the primary.  If we assumed that
all of the mass of the secondary is lost (an unphysical case since the
mass transfer would end before this could happen, but this provides an
extreme upper limit) then while the secondary would have lost
1$M_\odot$ the primary would have only accreted 0.3$M_\odot$ resulting
in a maximum primary mass of 1.7 $M_\odot$.  If we were to assume that
mass transfer would stop when the secondary decreased to a
mass of 0.3$M_\odot$ then the secondary would have lost 0.7$M_\odot$
and only 0.21$M_\odot$ (or 30\% of 0.7$M_\odot$) would have been accreted by the primary
resulting in a mass of 1.61$M_\odot$.  We have examined 21 curves like
those in Figure~\ref{fig:sim3}, we measured and plotted their peak
value and the full-width-half-maximum (FWHM) in
Figure~\ref{fig:AccRate}.  The error bars denote the FWHM of the
curves, the plotted point is the peak value for each curve, and the
mass of J1903+0327 is shown as a dashed line.  Examination of
Figure~\ref{fig:AccRate} shows that given the observed mass and the
assumptions we used in preparing the simulated systems, J1903+0327's progenitor
system would have most likely had an $F_{acc}$ value between between
0.35 and 0.5, with the peak value of 0.4 most closly maching the
observed mass.

It should be noted however, not all of the barionic mass transfered results in an equivalent 
increase in gravitational mass of the primary since $M_{accrete}$ = $\Delta M_{grav} + \Delta E_{binding}/c^2$
\citep{2011MNRAS.413L..47B}, where $M_{accrete}$ is the mass accreted from the secondary, $\Delta M_{grav}$
is the change in gravitational mass of the primary, and $\Delta E_{binding}$ is the binding energy of 
the system.  We find that for the masses being transferred in our simulations
the effect of using $M_{accrete} = \Delta M_{grav}$ is less than the uncertainty in the final results.

\begin{figure}
 \includegraphics[angle=270, width=\linewidth]{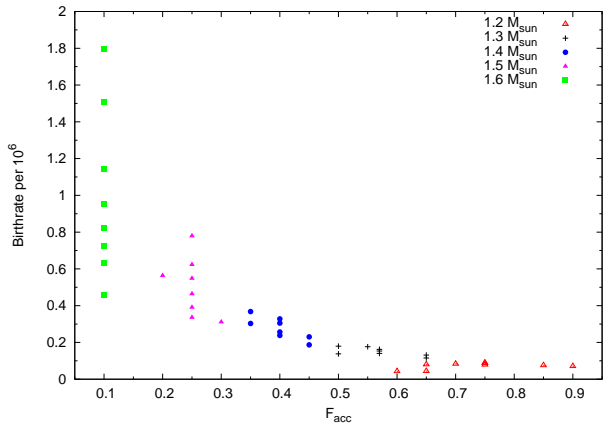}
  \caption{The number of systems per millon simulations with a final primary mass of 1.667 $M_\odot$
  (the observed mass of J1903+0327) as a function of the fraction of accreted mass, for different
  initial primary masses (shown in the key).  
  \label{fig:BirthRate} }
\end{figure}

Finally, we preform the same analysis that produced Figure~\ref{fig:sim3} but use an initial primary
mass of 1.2, 1.3, 1.4 (as used in all of the previous simulations), 1.5, and 1.6 $M_\odot$.  These simulations 
were preformed for eight inner semi-major axes (10, 20, 30, 40, 50, 60, 70, and 100 $R_\odot$) at the start of mass transfer.  
The $F_{acc}$ value with the peak number of occurrences closest to the observed mass of J1903+0327
(1.667 $M_\odot$) was recorded, as was the number of occurrences at that peak; these values were plotted in Figure~\ref{fig:BirthRate}.  
Upon examining Figure~\ref{fig:BirthRate} we find that as the initial mass of the primary increases the most likely $F_{acc}$ value and its domain decrease.  
To understand these results we recall that as the initial mass of the primary increases the amount of mass needed to reach the observed mass of J1903+0327
is decreased.  So, for example, if the initial mass of the progenitor of J1903+0327's primary (before it began to accrete material from the secondary)
was 1.6 $M_\odot$ it would only need to accrete 0.067 $M_\odot$ before the system reached the observed mass.  
A very small $F_{acc}$ value can result in the transfer of such a small amount of material allowing the $F_{acc}$ to stay low;
with a lager $F_{acc}$ value the system will often reach a final primary mass greater than 1.667 $M_\odot$ thus limiting the domain.  
Whereas if the initial primary mass was 1.2 $M_\odot$, an $F_{acc}$ value of 0.1 would never allow for enough mass to be transfered,
but there are a large range of $F_{acc}$ values that can allow for that amount of mass transfer that would not quickly overshoot 
the observed mass.  This assumes, as we have in all of the simulations, that the mass transfer is stable as long as the triple is dynamically stable.
We find that for an initial primary mass of 1.4 $M_\odot$, the value used in all previous simulations, the peak $F_{acc}$ value
is not sensitive to the semi-major axis at the beginning of the mass transfer; the $F_{acc}$ value ranges between 0.35 and 0.45 which lies within 
our expected range of 0.35 to 0.5 found above from Figure~\ref{fig:AccRate}.  

\section{Conclusion}

We have examined the effect of an asymmetric supernova (SN) on a
hierarchical multiple star system and considered how it can be modeled
by applying the effective binary method. This is done by recursively
replacing the inner binary by an effective star at the center of mass
of that binary. The effective star experiences an effective SN with
the effects of sudden mass loss, an instantaneous translation and an
effective kick velocity, i.e. the systemic velocity of the inner
binary. We have coded the equations in this paper in a small python
script, which is publicly available\footnote{The source code is
  publicly available at {\tt
    http://castle.strw.leidenuniv.nl/software.html}.}
    
We point out that the effective SN is different from a physical SN in that
for a physical SN themass is lost from the position of the physical star,
whereas for an effective SN the mass is lost from the effective star.  
The off-center mass loss in an effective SN becomes
important only if the shell impact on the companion(s) is considered,
and otherwise causes no difference between a real and effective SN
calculation.  Furthermore, we calculated the runaway velocities for
dissociating binaries and effective binaries. We subsequently
demonstrated how calculating the effect of a SN on a multiple can be
generalized to multiples in which a star other than the primary is
undergoing the SN.

We used this method to examine the case for J1903+0327 forming from a
hierarchical triple.  We assume initial masses of 10, 1.0, and 0.9
$M_\odot$ for the primary, secondary, and tertiary respectively, as
well as an inner semi-major axis of 200 $R_\odot$.  We find that if
J1903+0367 was to form through such a mechanism it would be most
likely to have a very low SN kick velocity so that it would remain
stable after the SN, and a large inner semi-major axis after the CE
phase to increase the likelihood that the triple would become unstable
once the NS/MSP reached a mass of 1.667 $M_\odot$
\citep{2011MNRAS.412.2763F}.  We also find that, given our
assumptions, the transfer efficiency, $F_{acc}$, for J1903+0327 would
have likely been between 0.35 and 0.5.

\section{Acknowledgements}\label{sec:Acknowledgements}
We thank Elena Maria Rossi for her useful suggestions.  We would also like to thank the referee
for his helpful comments which have greatly improved this paper.  This work
was supported by the Netherlands Research School for Astronomy (NOVA) and
the Netherlands Research Council NWO [grants VICI (639.073.803) and AMUSE
(614.061.608)].

\bibliographystyle{mn2e}
\bibliography{triplebib}



\label{lastpage}

\end{document}